\documentclass{aastex}
\usepackage{amssymb}

\begin{document}

\begin{center}
{\LARGE Multicomponent theory of buoyancy instabilities in astrophysical
plasma objects: The case of magnetic field perpendicular to gravity}
\end{center}

{\LARGE \bigskip }

\begin{center}
{\large Anatoly K. Nekrasov}

Institute of Physics of the Earth, Russian Academy of Sciences, 123995
Moscow, Russia

anatoli.nekrassov@t-online.de

\bigskip

{\large and}

\bigskip

{\large Mohsen Shadmehri}

Department of Mathematical Physics, National University of Ireland Maynooth,
Maynooth, Co. Kildare, Ireland\\[0pt]
NORDITA, AlbaNova University Center, Roslagstullsbacken 23, SE-10691
Stockholm, Sweden\\[0pt]

mshadmehri@thphys.nuim.ie

\bigskip

{\large ABSTRACT}
\end{center}

We develop a general theory of buoyancy instabilities in the electron-ion
plasma with the electron heat flux based not upon MHD equations, but using a
multicomponent plasma approach in which the momentum equation is solved for
each species. We investigate the geometry in which the background magnetic
field is perpendicular to the gravity and stratification. General
expressions for the perturbed velocities are given without any
simplifications. Collisions between electrons and ions are taken into
account in the momentum equations in a general form, permitting us to
consider both weakly and strongly collisional objects. However, the electron
heat flux is assumed to be directed along the magnetic field that implies a
weakly collisional case. Using simplifications justified for an
investigation of buoyancy instabilities with the electron thermal flux, we
derive simple dispersion relations both for collisionless and collisional
cases for arbitrary directions of the wave vector. The collisionless
dispersion relation considerably differs from that obtained in the MHD
framework and is similar to the Schwarzschild's criterion. This difference
is connected with simplified assumptions used in the MHD analysis of
buoyancy instabilities and with the role of the longitudinal electric field
perturbation which is not captured by the ideal MHD equations. The results
obtained can be applied to clusters of galaxies and other astrophysical
objects.

\bigskip \textit{Subject headings: }convection - instabilities - magnetic
fields - plasmas - waves

\bigskip

\section{INTRODUCTION}

\bigskip

Physical processes taking place in astrophysical objects are defined by the
physical parameters of the latter. In many cases, evolution of these
parameters can lead to instabilities influencing the dynamics of these
objects. Convective or buoyancy instabilities arising as a result of
stratification of astrophysical objects are among those instabilities that
may operate under different conditions from stellar interiors (e.g.,
Schwarzschild 1958), accretion disks (Balbus 2000, 2001), and neutron stars
(Chang \& Quataert 2009) to hot accretion flows (e.g., Narayan et al. 2000,
2002) and even galaxy clusters and intracluster medium (ICM) (e.g., Quataert
2008; Sharma et al. 2009; Ren et al. 2009). Analogous instabilities also
exist in the neutral atmosphere of the Earth and ocean (Gossard \& Hooke
1975; Pedlosky 1982). It is believed that convective instabilities have a
vital role not only in transporting energy but in driving turbulence in many
astrophysical systems. When the entropy increases in the direction of
gravity, a thermally stratified fluid becomes buoyantly unstable according
to the Schwarzschild criterion (Schwarzschild 1958). This well-known
condition is successfully applied in the theoretical modeling of stellar
structure and is valid irrespective of existence of the magnetic field. But
a stellar fluid is a strong collisional system. At the same time, there are
systems in which plasma is tenuous and hot such as, for example, ICM
(Sarazin 1988). Majority of the mass of a galaxy cluster is in the dark
matter. However, around 1/6 of its mass consists of hot, magnetized, and low
density plasma known as ICM. The ICM is classified as a weakly collisional
plasma with the electron number density $n_{e}\simeq 10^{-2}-10^{-1}$ cm$%
^{-3}$, the electron temperature $T_{e}$ of the order of a few keV (e.g.,
Fabian et al. 2006; Sanders et al. 2010), and the magnetic field strength $%
B\simeq 0.1-10$ $\mu $G (Carilli \& Taylor 2002). Thus in ICM, the mean free
path of ions and electrons is much larger than their Larmor radius. Cosmic
rays play also an important role in the physics of ISM. Recent studies show
that centrally concentrated cosmic rays have a destabilizing effect on the
convection in ICM (Chandran \& Dennis 2006; Rasera and Chandran 2008).

In the recent past, one has included the anisotropic heat flux in weakly
collisional plasmas and obtained additional instabilities. These
instabilities have been shown to arise when the temperature increases in the
direction of gravity, if the background thermal flux is absent (the
magnetothermal instability (MTI)) (Balbus 2000, 2001), and when the
temperature decreases along the gravity at the presence of the latter (the
heat buoyancy instability (HBI)) (Quataert 2008). The growth rates of MTI
and HBI are of the same order of magnitude as the growth rates without heat
flux.

Previous theoretical models applied for study of buoyancy instabilities are
based on the ideal magnetohydrodynamic (MHD) equations (Balbus 2000, 2001;
Quataert 2008; Chang \& Quataert 2009; Ren et al. 2009). However, the ideal
MHD does not capture some important effects. One such effect is the nonzero
longitudinal electric field perturbation along the magnetic field. We show
here that the contribution of currents due to this small (at the present
case) field to the dispersion relation can be of the same order of magnitude
as that due to the electric field components transverse to the magnetic
field. Besides, the MHD equations can not take into account the existence of
various charged and neutral species with the different masses and electric
charges and collisions of different species with each other. On the
contrary, the plasma $\mathbf{E}$-approach deals with dynamical equations
for each species. Starting with Faraday's and Ampere's laws, one obtains
equations for the electric field components. Such an approach allows us to
follow the movement and change of parameters of each species separately and
to obtain rigorous conditions of consideration and corresponding physical
consequences in specific cases. This approach permits us to include various
species of ions and dust grains having different charges and masses. In this
way, streaming instabilities of rotating multicomponent objects with
different background velocities of species (accretion disks, molecular
clouds and so on) have been investigated by Nekrasov (2008, 2009 a, b, c,
d). It has been shown that these instabilities have the growth rates much
larger than that of the magnetorotational instability (Balbus 1991). In some
cases, the standard methods used in MHD leads to conclusions that
significantly differ from those obtained by the method using the electric
field perturbations. One of such examples concerning the contribution of
collisions in the electron-ion plasma has been shown in (Nekrasov 2009 c).

A study of buoyancy instabilities with the electron heat flux by the
multicomponent $\mathbf{E}$-approach has been performed by Nekrasov and
Shadmehri (2010). A geometry has been considered in which the gravity,
stratification, and background magnetic field are all directed along the one
axis. Solution of the dispersion relation has shown that the thermal flux
has a stabilizing effect. The same problem solved by the ideal MHD leads to
another result (Quataert 2008).

In this paper, we apply the multicomponent $\mathbf{E}$-approach to study
buoyancy instabilities in magnetized nonuniform electron-ion astrophysical
plasmas in which the background magnetic field is perpendicular to the
gravitational field. The inhomogeneity is assumed to be directed along the
gravity. We include collisions between electrons and ions and electron
thermal conductivity. The dynamical frequency and the wave phase velocity
along the magnetic field are assumed much less than the ion cyclotron
frequency and electron thermal velocity, respectively. In this case, the
electrons can be regarded as inertialess. The pressure anisotropy of species
is not taken into account. Thus we exclude the pressure-anisotropy-driven
firehose (e.g., Vedenov \& Sagdeev 1958; Kennel \& Sagdeev 1967 a, b;
Schekochihin et al. 2008) and mirror (e.g., Vedenov \& Sagdeev 1959,
Southwood \&\textbf{\ }Kivelson 1993; Califano et al. 2008) mode
instabilities. We also neglect the viscous and finite Larmor radius effects.
The corresponding conditions can be easily obtained from momentum equations
given, for example, in\textbf{\ (}Nekrasov 2009 c, d). Solutions for
perturbed velocities of species are found in a general form. Using
conditions suitable for buoyancy perturbations, we obtain simplified
expressions for perturbed values. The final dispersion relation is derived
for the case in which the anisotropic thermal flux is important. The growth
rates are found for collisionless as well as collisional cases. Solutions of
the dispersion relation are discussed.

The paper is organized as follows. In Section 2, the fundamental equations
are given. The equilibrium state is considered in Section 3. The perturbed
ion velocity, number density, and thermal pressure are obtained in Section
4. In Section 5, we consider the perturbed velocity and temperature for
electrons. Perturbed current components are calculated in Section 6. In
Section 7, we consider conditions to simplify the contribution from
collisions. The dispersion relation is derived in Section 8. In Section 9,
the main points of our analysis are discussed. Implications of results
obtained for galaxy clusters are considered in Section 10. Concluding
remarks are given in Section 11.

\bigskip

\section{BASIC EQUATIONS}

\bigskip

We start with the following equations for ions:

\begin{equation}
\frac{\partial \mathbf{v}_{i}}{\partial t}+\mathbf{v}_{i}\cdot \mathbf{%
\nabla v}_{i}\mathbf{=-}\frac{\mathbf{\nabla }p_{i}}{m_{i}n_{i}}+\mathbf{g+}%
\frac{q_{i}}{m_{i}}\mathbf{E}+\frac{q_{i}}{m_{i}c}\mathbf{v}_{i}\times 
\mathbf{B}-\nu _{ie}\left( \mathbf{v}_{i}-\mathbf{v}_{e}\right) ,
\end{equation}%
the momentum equation, 
\begin{equation}
\frac{\partial n_{i}}{\partial t}+\mathbf{\nabla }\cdot n_{i}\mathbf{v}%
_{i}=0,
\end{equation}%
the continuity equation, and 
\begin{equation}
\frac{\partial p_{i}}{\partial t}+\mathbf{v}_{i}\cdot \mathbf{\nabla }%
p_{i}+\gamma p_{i}\mathbf{\nabla }\cdot \mathbf{v}_{i}=0,
\end{equation}%
the pressure equation. The corresponding equations for the inertialess
electrons are:

\begin{equation}
\mathbf{0=-}\frac{\mathbf{\nabla }p_{e}}{n_{e}}+q_{e}\mathbf{E}+\frac{q_{e}}{%
c}\mathbf{v}_{e}\times \mathbf{B}-m_{e}\nu _{ei}\left( \mathbf{v}_{e}-%
\mathbf{v}_{i}\right) ,
\end{equation}%
\begin{equation}
\frac{\partial n_{e}}{\partial t}+\mathbf{\nabla }\cdot n_{e}\mathbf{v}%
_{e}=0,
\end{equation}%
\begin{equation}
\frac{\partial p_{e}}{\partial t}+\mathbf{v}_{e}\cdot \mathbf{\nabla }%
p_{e}+\gamma p_{e}\mathbf{\nabla }\cdot \mathbf{v}_{e}=-\left( \gamma
-1\right) \mathbf{\nabla \cdot q}_{e},
\end{equation}%
\begin{equation}
\frac{\partial T_{e}}{\partial t}+\mathbf{v}_{e}\cdot \mathbf{\nabla }%
T_{e}+\left( \gamma -1\right) T_{e}\mathbf{\nabla }\cdot \mathbf{v}%
_{e}=-\left( \gamma -1\right) \frac{1}{n_{e}}\mathbf{\nabla \cdot q}_{e},
\end{equation}%
where the last equation is the temperature equation and $\mathbf{q}_{e}$ is
the electron heat flux (Braginskii 1965). In Equations (1)-(7), $q_{j}$ and $%
m_{j}$ are the charge and mass of species $j=i,e$, $\mathbf{v}_{j}$ is the
hydrodynamic velocity, $n_{j}$ is the number density, $p_{j}=n_{j}T_{j}$ is
the thermal pressure, $T_{j}$ is the temperature, $\nu _{ie}$ ($\nu _{ei}$)
is the collision frequency of ions (electrons) with electrons (ions), $%
\mathbf{g}$ is the gravity, $\mathbf{E}$\textbf{\ }and $\mathbf{B}$ are the
electric and magnetic fields, $c$ is the speed of light in vacuum, and $%
\gamma $ is the adiabatic constant. As for the electron heat flux, we
consider the case of weakly collisional plasma for which the electron
cyclotron frequency $\omega _{ce}=q_{e}B/m_{e}c$ is much larger than the
electron-electron collision frequency $\nu _{ee}$, i.e. $\omega _{ce}\gg \nu
_{ee}$. In this case, the electron thermal flux is mainly directed along the
magnetic field,%
\begin{equation}
\mathbf{q}_{e}=-\chi _{e}\mathbf{b}\left( \mathbf{b\cdot \nabla }\right)
T_{e},
\end{equation}%
where $\chi _{e}$ is the electron thermal conductivity coefficient and $%
\mathbf{b=B/}B$ is the unit vector along the magnetic field. We only take
into account the electron thermal flux (8) because the ion thermal
conductivity is considerably smaller (Braginskii 1965). We also assume that
the thermal flux in equilibrium is negligible.

Electromagnetic equations are Faraday's 
\begin{equation}
\mathbf{\nabla \times E=-}\frac{1}{c}\frac{\partial \mathbf{B}}{\partial t},
\end{equation}%
and Ampere`s 
\begin{equation}
\mathbf{\nabla \times B=}\frac{4\pi }{c}\mathbf{j,}
\end{equation}%
laws, where $\mathbf{j=}\sum_{j}q_{j}n_{j}\mathbf{v}_{j}.$ We consider wave
processes with typical time-scales much larger than the time the light
spends to cover the wavelength of perturbations. In this case, one can
neglect the displacement current in Equation (10) that results in
quasi-neutrality for both the electromagnetic and purely electrostatic
perturbations. The magnetic field $\mathbf{B}$ includes the background
magnetic field $\mathbf{B}_{0}$, the magnetic field $\mathbf{B}_{0cur}$ of
the background electric current (when it presents), and the perturbed
magnetic field.

Our basic equations are the same as that in (Nekrasov \& Shadmehri 2010).
However, in the present case, the equilibrium state differs from the
equilibrium in (Nekrasov \& Shadmehri 2010), where magnetic field lines and
the gravity are parallel to each other. In the next section, we analyze the
equilibrium state in which the magnetic field is perpendicular to the
direction of gravity.

\bigskip

\section{\protect\bigskip EQUILIBRIUM}

\subsection{Background velocities}

At first, we consider the equilibrium state. In this paper, we study the
configuration in which the background magnetic field $\mathbf{B}_{0}$ is
parallel to the $z$-axis, and the gravity and stratification are parallel to
the $x$-axis. Let, for definiteness, $\mathbf{g}$ be $\mathbf{g=-x}g$, where 
$g>0$ and $\mathbf{x}$ is the unit vector along the $x$-direction. Then,
Equations (1) and (4) are the following:%
\begin{equation}
\mathbf{v}_{i0}\cdot \mathbf{\nabla v}_{i0}\mathbf{=}\mathbf{-}\frac{\mathbf{%
\nabla }p_{i0}}{m_{i}n_{i0}}+\mathbf{g+}\frac{q_{i}}{m_{i}}\mathbf{E}_{0}+%
\frac{q_{i}}{m_{i}c}\mathbf{v}_{i0}\times \mathbf{B}_{0}-\nu _{ie}^{0}\left( 
\mathbf{v}_{i0}-\mathbf{v}_{e0}\right) ,
\end{equation}%
\begin{equation}
\mathbf{0=-}\frac{\mathbf{\nabla }p_{e0}}{n_{e0}}+q_{e}\mathbf{E}_{0}+\frac{%
q_{e}}{c}\mathbf{v}_{e0}\times \mathbf{B}_{0}-m_{e}\nu _{ei}^{0}\left( 
\mathbf{v}_{e0}-\mathbf{v}_{i0}\right) ,
\end{equation}%
where the index $0$ denotes the background values. Our system is uniform
along the $y$- and $z$-axes. Therefore, we assume that $E_{0y}=E_{0z}=0$.
However, for generality, we keep $E_{0x}\neq 0$. We also assume that $%
v_{i0z}=v_{e0z}=0$. Then, from the $y$-components of Equations (11) and
(12), we obtain 
\begin{equation}
v_{i0x}\mathbf{=}v_{e0x}\mathbf{=}-\frac{\nu _{ie}^{0}}{\omega _{ci}}\left(
v_{i0y}-v_{e0y}\right)
\end{equation}%
provided that $\omega _{ci}=q_{i}B_{0}/m_{i}c\gg \partial v_{i0y}/\partial x$%
.\ In this case, the contribution of the ion inertia along the $y$-axis
produced by the shear of velocity $v_{i0y}$ along the $x$-axis can be
neglected in comparison with the $y$-component of the Lorentz force. From
the $x$-components of Equations (11) and (12), we also find under condition $%
\omega _{ci}v_{i0y}\gg \partial v_{i0x}^{2}/2\partial x$ 
\begin{equation}
v_{i0y}=v_{E}+v_{g}+v_{di}=-c\frac{E_{0x}}{B_{0}}+\frac{g}{\omega _{ci}}-%
\frac{g_{i}}{\omega _{ci}},
\end{equation}%
\begin{equation}
v_{e0y}=v_{E}+v_{de}\mathbf{=}-c\frac{E_{0x}}{B_{0}}+\frac{g_{e}}{\omega
_{ci}},
\end{equation}%
where%
\begin{equation}
g_{i}=\mathbf{-}\frac{1}{m_{i}n_{i0}}\frac{\partial p_{i0}}{\partial x}%
,g_{e}=\mathbf{-}\frac{1}{m_{i}n_{e0}}\frac{\partial p_{e0}}{\partial x}.
\end{equation}

The velocities $v_{E}$, $v_{g}$, and $v_{di,e}$ are the electric,
gravitational, and diamagnetic drifts, respectively. We assume that $%
q_{i}=-q_{e}$. If values $g_{i}$ and $g_{e}$ are inhomogeneous and $%
g_{i}\sim g_{e}$, then conditions given above to justify solutions (13) and
(14) can be written in the following form: 
\begin{equation}
1\gg \left( 1+\frac{\nu _{ie}^{02}}{\omega _{ci}^{2}}\right) \frac{\rho
_{i}^{2}}{L^{2}},
\end{equation}%
where $\rho _{i}=v_{Ti}/\omega _{ci}$ ($v_{Ti}$ is the ion thermal velocity)
is the ion Larmor radius and $L^{-1}=\partial g_{i,e}/g_{i,e}\partial x$ (an
uncertain contribution of the electric drift is not taken into account in
condition [17]). We note that there are no restrictions on the relation
between $\omega _{ci}$ and $\nu _{ie}^{0}$ in inequality (17), i.e. the case 
$\nu _{ie}^{0}>\omega _{ci}$ can also take place.

\bigskip

\subsection{Continuity and pressure equations}

Having the background velocities, we can now consider the ion continuity
equation (2). In equilibrium, we have%
\begin{equation}
\frac{\partial n_{i0}}{\partial t}+\frac{\partial }{\partial x}%
n_{i0}v_{i0x}=0.
\end{equation}%
This equation is nonstationary because of the $x$-dependence of $v_{i0x}$.
The typical time $t_{0}$ of the density evolution is $t_{0}\sim 1/\nu
_{ie}^{0}\kappa ^{2}\rho _{i}^{2}$, where $\kappa =\partial
n_{0}/n_{0}\partial x$ ($n_{i0}=n_{e0}=n_{0}$). When studying the linear
perturbations, we will not take into account that the medium is
nonstationary. This means that we will consider perturbations which develop
much faster than $t_{0}$. Thus, we can safely assume that the medium is
stationary. The same also relates to the ion pressure equation and
corresponding equations for electrons.

\bigskip

\section{LINEAR\ ION\ PERTURBATIONS}

\bigskip

\subsection{Equation for the perturbed ion velocity}

Let us write the equation of motion (1) for ions in the linear approximation,%
\begin{eqnarray}
\frac{\partial \mathbf{v}_{i1}}{\partial t}+\mathbf{v}_{i0}\cdot \mathbf{%
\nabla v}_{i1}+\mathbf{v}_{i1}\cdot \mathbf{\nabla v}_{i0} &\mathbf{=}&-%
\frac{\mathbf{\nabla }p_{i1}}{m_{i}n_{i0}}+\frac{\mathbf{\nabla }p_{i0}}{%
m_{i}n_{i0}}\frac{n_{i1}}{n_{i0}}\mathbf{+}\frac{q_{i}}{m_{i}}\mathbf{E}_{1}+%
\frac{q_{i}}{m_{i}c}\mathbf{v}_{i1}\times \mathbf{B}_{0} \\
&&+\frac{q_{i}}{m_{i}c}\mathbf{v}_{i0}\times \mathbf{B}_{1}-\nu
_{ie}^{0}\left( \mathbf{v}_{i1}-\mathbf{v}_{e1}\right) -\nu _{ie}^{1}\left( 
\mathbf{v}_{i0}-\mathbf{v}_{e0}\right) ,  \nonumber
\end{eqnarray}%
where the index $1$ denotes the values of the first order of magnitude. The
diamagnetic drift in the stationary velocity $\mathbf{v}_{i,e0}$ is not a
real velocity (e.g. Nishikawa \& Wakatani 1990). However, this drift must be
taken into account in the hydrodynamical equations evoking drift waves (e.g.
Vranje\v{s} et al. 2003 for the same initial state). The last collisional
term in Equation (19) appears as a result of perturbation of collision
frequency due to the density and temperature perturbations: $\nu
_{ie}^{1}/\nu _{ie}^{0}=$ $n_{e1}/n_{e0}-3T_{e1}/2T_{e0}$. This effect due
to the density perturbation has been involved by Nekrasov (e.g. 2009 b).

Below, we do not include the shear of the background velocity on the left
hand-side of Equation (19) (the same also relates to the continuity and
pressure equations). It is possible, if 
\begin{equation}
D_{ti}=\frac{\partial }{\partial t}+\mathbf{v}_{i0}\cdot \mathbf{\nabla }\gg 
\frac{\partial v_{i0x}}{\partial x};\frac{\partial v_{i0y}}{\partial x}.
\end{equation}%
The first inequality (20), $D_{ti}\gg \partial v_{i0x}/\partial x$,
coincides with the condition of medium stationarity (see Eq. [18]). The
second inequality, $D_{ti}\gg \partial v_{i0y}/\partial x$, is satisfied
automatically if $v_{i0y}=\mathrm{const}$. When $v_{i0y}\neq \mathrm{const}$%
, this condition can be written in the form $D_{ti}\gg \omega _{ci}\rho
_{i}^{2}/L^{2}$. We further introduce the following notations:%
\begin{eqnarray}
\mathbf{F}_{i1} &=&\frac{q_{i}}{m_{i}}\mathbf{E}_{1}-\nu _{ie}^{0}\left( 
\mathbf{v}_{i1}-\mathbf{v}_{e1}\right) , \\
\mathbf{G}_{i1} &=&\mathbf{F}_{i1}+\frac{q_{i}}{m_{i}c}\mathbf{v}_{i0}\times 
\mathbf{B}_{1},  \nonumber \\
\mathbf{C}_{i1} &=&\nu _{ie}^{1}\left( \mathbf{v}_{i0}-\mathbf{v}%
_{e0}\right) .  \nonumber
\end{eqnarray}%
Then Equation (19) takes the form%
\begin{equation}
D_{ti}\mathbf{v}_{i1}\mathbf{=}-\frac{\mathbf{\nabla }p_{i1}}{m_{i}n_{i0}}+%
\frac{\mathbf{\nabla }p_{i0}}{m_{i}n_{i0}}\frac{n_{i1}}{n_{i0}}\mathbf{+G}%
_{i1}-\mathbf{C}_{i1}+\frac{q_{i}}{m_{i}c}\mathbf{v}_{i1}\times \mathbf{B}%
_{0}.
\end{equation}

\bigskip

\subsection{Perturbed ion continuity and pressure equations}

In the linear approximation, the ion continuity (Eq. [2]) and pressure (Eq.
[3]) equations are given by

\begin{equation}
D_{ti}n_{i1}+v_{i1x}\frac{\partial n_{i0}}{\partial x}+n_{i0}\mathbf{\nabla }%
\cdot \mathbf{v}_{i1}=0,
\end{equation}

\begin{equation}
D_{ti}p_{i1}+v_{i1x}\frac{\partial p_{i0}}{\partial x}+\gamma p_{i0}\mathbf{%
\nabla }\cdot \mathbf{v}_{i1}=0.
\end{equation}%
Equations (22)-(24) are used to find perturbed ion velocity components.
These calculations are given in the Appendix A. For simplicity, we adopt
that $g_{i,e}=\mathrm{const}$ (see Eq. [16]). This is true when the gravity
is directed along the background magnetic field (Nekrasov \& Shadmehri
2010). This case is also satisfied, if we impose the condition $v_{i,e1y}=0$
(see Eqs. [14] and [15]). However, in a general case $g_{i,e}\neq \mathrm{%
const}$.

\bigskip

\section{LINEAR\ ELECTRON\ PERTURBATIONS}

\subsection{Equation for the perturbed electron velocity}

Consider Equation (4) in the linear approximation

\begin{equation}
\mathbf{0=-}\frac{\mathbf{\nabla }p_{e1}}{n_{e0}}+\frac{\mathbf{\nabla }%
p_{e0}}{n_{e0}}\frac{n_{e1}}{n_{e0}}+q_{e}\mathbf{E}_{1}+\frac{q_{e}}{c}%
\mathbf{v}_{e1}\times \mathbf{B}_{0}\mathbf{+}\frac{q_{e}}{c}\mathbf{v}%
_{e0}\times \mathbf{B}_{1}-m_{e}\nu _{ei}^{0}\left( \mathbf{v}_{e1}-\mathbf{v%
}_{i1}\right) -m_{e}\nu _{ei}^{1}\left( \mathbf{v}_{e0}-\mathbf{v}%
_{i0}\right) .
\end{equation}%
Here $\nu _{ei}^{1}/\nu _{ei}^{0}=$ $n_{i1}/n_{i0}-3T_{e1}/2T_{e0}$. We
introduce the following notations:%
\begin{eqnarray}
\mathbf{F}_{e1} &=&q_{e}\mathbf{E}_{1}-m_{e}\nu _{ei}^{0}\left( \mathbf{v}%
_{e1}-\mathbf{v}_{i1}\right) , \\
\mathbf{G}_{e1} &=&\mathbf{F}_{e1}\mathbf{+}\frac{q_{e}}{c}\mathbf{v}%
_{e0}\times \mathbf{B}_{1},  \nonumber \\
\mathbf{C}_{e1} &=&m_{e}\nu _{ei}^{1}\left( \mathbf{v}_{e0}-\mathbf{v}%
_{i0}\right) .  \nonumber
\end{eqnarray}%
Then Equation (25) takes the form%
\begin{equation}
\mathbf{0=-}\frac{\mathbf{\nabla }p_{e1}}{n_{e0}}+\frac{\mathbf{\nabla }%
p_{e0}}{n_{e0}}\frac{n_{e1}}{n_{e0}}+\mathbf{G}_{e1}-\mathbf{C}_{e1}+\frac{%
q_{e}}{c}\mathbf{v}_{e1}\times \mathbf{B}_{0}.
\end{equation}

\bigskip

\subsection{Perturbed electron continuity, pressure, and temperature
equations}

The continuity equation (5) in the linear approximation is given by 
\begin{equation}
D_{te}n_{e1}+v_{e1x}\frac{\partial n_{e0}}{\partial x}+n_{e0}\mathbf{\nabla }%
\cdot \mathbf{v}_{e1}=0,
\end{equation}%
where $D_{te}=\partial /\partial t+\mathbf{v}_{e0}\cdot \mathbf{\nabla }$.
Here, we also neglect the contribution of equilibrium velocity
inhomogeneity, i.e. we assume that $D_{te}\gg \partial v_{e0x}/\partial x$.

The linear electron pressure equation (6) is the following: 
\begin{equation}
D_{te}p_{e1}+v_{e1x}\frac{\partial p_{e0}}{\partial x}+\gamma p_{e0}\mathbf{%
\nabla }\cdot \mathbf{v}_{e1}+\left( \gamma -1\right) \mathbf{\nabla \cdot q}%
_{e1}=0,
\end{equation}%
where $\mathbf{q}_{e1}$ is the linear electron thermal flux.

The perturbed electron temperature equation (7) has the form

\begin{equation}
D_{te}T_{e1}+v_{e1x}\frac{\partial T_{e0}}{\partial x}+\left( \gamma
-1\right) T_{e0}\mathbf{\nabla }\cdot \mathbf{v}_{e1}=-\left( \gamma
-1\right) \frac{1}{n_{e0}}\mathbf{\nabla \cdot q}_{e1}.
\end{equation}

\subsection{Perturbed electron thermal flux}

Expression for the thermal flux (8) in the linear approximation is 
\begin{equation}
\mathbf{q}_{e1}=-\chi _{e0}\mathbf{b}_{0}\left( \mathbf{b}_{1}\mathbf{\cdot
\nabla }\right) T_{e0}-\chi _{e0}\mathbf{b}_{0}\frac{\partial T_{e1}}{%
\partial z},
\end{equation}%
where 
\[
\mathbf{b}_{1}=\frac{\mathbf{B}_{1}}{B_{0}}-\mathbf{b}_{0}\frac{B_{1z}}{B_{0}%
}. 
\]%
Thus from Equation (31), we have%
\begin{eqnarray}
q_{e1x,y} &=&0, \\
q_{e1z} &=&-\chi _{e0}\frac{\partial T_{e0}}{\partial x}\frac{B_{1x}}{B_{0}}%
-\chi _{e0}\frac{\partial T_{e1}}{\partial z}.  \nonumber
\end{eqnarray}%
Using Equation (32), we find the value $\mathbf{\nabla \cdot q}_{e1}$ as
follows 
\begin{equation}
\mathbf{\nabla \cdot q}_{e1}=-\chi _{e0}\frac{\partial T_{e0}}{\partial x}%
\frac{1}{B_{0}}\frac{\partial B_{1x}}{\partial z}-\chi _{e0}\frac{\partial
^{2}T_{e1}}{\partial z^{2}}.
\end{equation}%
The first term on the right hand-side of Equation (33) can be considered as
a source of the heat, and the second term describes the thermal diffusion.
This expression for $\mathbf{\nabla \cdot q}_{e1}$ is analogous to the one
arising in the case in which the stratification is directed along the
background magnetic field and there exists an anisotropic background thermal
flux (see Nekrasov \& Shadmehri 2010, Eq. [49]).

\bigskip\ 

\subsection{Perturbed electron temperature and pressure equations with
the thermal flux}

Equation (30), taking into account expression (33), can be written in the
form%
\begin{equation}
\left( D_{te}+\Omega \right) T_{e1}=-v_{e1x}\frac{\partial T_{e0}}{\partial x%
}-\left( \gamma -1\right) T_{e0}\mathbf{\nabla }\cdot \mathbf{v}_{e1}-\frac{%
\partial T_{e0}}{\partial x}\frac{1}{B_{0}}\Omega \left( \frac{\partial }{%
\partial z}\right) ^{-1}B_{1x},
\end{equation}%
where the operator

\begin{equation}
\Omega =-\left( \gamma -1\right) \frac{1}{n_{e0}}\chi _{e0}\frac{\partial
^{2}}{\partial z^{2}}
\end{equation}%
is introduced.

Equation (34) permits us to define the value $\mathbf{\nabla \cdot q}_{e1}$.
Substituting this value in Equation (29), we obtain the desirable equation
for the perturbed electron pressure%
\begin{eqnarray}
0 &=&D_{te}p_{e1}+\left[ \frac{\partial p_{e0}}{\partial x}-n_{e0}\frac{%
\Omega }{\left( D_{te}+\Omega \right) }\frac{\partial T_{e0}}{\partial x}%
\right] v_{e1x}+p_{e0}\frac{\gamma D_{te}+\Omega }{\left( D_{te}+\Omega
\right) }\mathbf{\nabla }\cdot \mathbf{v}_{e1} \\
&&+n_{e0}\frac{\partial T_{e0}}{\partial x}\frac{1}{B_{0}}\frac{D_{te}\Omega 
}{\left( D_{te}+\Omega \right) }\left( \frac{\partial }{\partial z}\right)
^{-1}B_{1x}.  \nonumber
\end{eqnarray}

Equations (27), (28), and (36) are used in the Appendix B for determining
the velocity $\mathbf{v}_{e1}$. For convenience and compactness of all the
equations and expressions, we formally apply the method of division by time
and spatial derivatives (see Eqs. [34], [36], and below). This permits us
not to use the Fourier transformation at the initial stage of calculations,
but rather gives a possibility to obtain the spatiotemporal differential
equations for variables under study (for example, for numerical
calculations). When the dispersion relation is derived in Section 8, all
derivatives will be changed by their\ Fourier-images.

\bigskip

\section{CURRENT $\mathbf{j}_{1}$}

\bigskip

Expressions for $\mathbf{v}_{i1}$ and $\mathbf{v}_{e1}$ obtained in the
Appendices A and B have a sufficiently general form. In particular, one can
take into account the background velocities of species that results in
electromagnetic streaming instabilities. Some studies of such instabilities
in astrophysical objects are given in Nekrasov (2007; 2008; 2009 a,b,c,d).
However, in the present paper, we do not consider this effect, assuming that%
\begin{eqnarray}
D_{t} &\gg &\mathbf{v}_{i,e0}\cdot \mathbf{\nabla ,} \\
D_{t}\mathbf{E}_{1} &\gg &\mathbf{\nabla v}_{i,e0}\cdot \mathbf{E}_{1}, 
\nonumber
\end{eqnarray}%
where $D_{t}=\partial /\partial t$. Thus, $D_{ti,e}\simeq D_{t}$. The second
inequality (37) is obtained from condition $c\mathbf{E}_{1}\gg \mathbf{v}%
_{i,e0}\times \mathbf{B}_{1}$ (see Eqs. [21] and [26]) and by using Equation
(9). Under conditions (37), one can also neglect the contribution of $%
\mathbf{C}_{i,e1}$ in Equations (22) and (27). We further consider
perturbations with the dynamical frequency much smaller than the ion sound
frequency. Keeping the spatiotemporal derivatives, this means that $%
c_{si}^{2}\partial ^{2}/\partial z^{2}\gg D_{t}^{2}$. We also assume a local
approximation when the perturbation wavelength in the direction of
inhomogeneity is smaller than the inhomogeneity scale, i.e.\textbf{\ }$%
\partial /\partial x\gg g_{i,e}/c_{si,e}^{2}$\textbf{. }Below, we find
components of the perturbed current $\mathbf{j}_{1}$. For simplicity, we
omit the index $0$ by $\nu _{ie}^{0}$.

\bigskip

\subsection{Current $j_{1x}$}

From Equation (A15) and under conditions given above, we obtain 
\begin{eqnarray}
v_{i1x} &\mathbf{=}&\frac{D_{t}}{\omega _{ci}^{2}}\left( \frac{\partial ^{2}%
}{\partial y^{2}}+\frac{\partial ^{2}}{\partial z^{2}}\right) \left( \frac{%
\partial }{\partial z}\right) ^{-2}F_{i1x}\mathbf{+}\frac{1}{\omega _{ci}}%
F_{i1y}-\frac{D_{ti}}{\omega _{ci}^{2}}\frac{\partial ^{2}}{\partial
x\partial y}\left( \frac{\partial }{\partial z}\right) ^{-2}F_{i1y} \\
&&-\frac{1}{\omega _{ci}}\frac{\partial }{\partial y}\left( \frac{\partial }{%
\partial z}\right) ^{-1}F_{i1z}-\frac{D_{t}}{\omega _{ci}^{2}}\left[ \frac{%
\omega _{ci}D_{t}}{c_{si}^{2}}\frac{\partial }{\partial y}\left( \frac{%
\partial }{\partial z}\right) ^{-2}+\frac{\partial }{\partial x}\right]
\left( \frac{\partial }{\partial z}\right) ^{-1}F_{i1z},  \nonumber
\end{eqnarray}%
where the additional condition $\partial /\partial x\gg \left(
g_{i}/c_{si}^{2}\right) \left( \partial /\partial y\right) ^{2}\left(
\partial /\partial z\right) ^{-2}$ has been assumed. From Equation (B4), we
have 
\begin{equation}
v_{e1x}\mathbf{=}\frac{1}{m_{e}\omega _{ce}}F_{e1y}-\frac{1}{m_{e}\omega
_{ce}}\frac{\partial }{\partial y}\left( \frac{\partial }{\partial z}\right)
^{-1}F_{e1z}.
\end{equation}%
Expression (38) coincides with expression (39) in the limit $%
m_{i}\rightarrow 0$ at substitution $i\leftrightarrow e$ and $m_{i}\mathbf{F}%
_{i1}\rightarrow \mathbf{F}_{e1}$.

Using expressions (38) and (39), we can find the current $j_{1x}$. It is
convenient to find the value $4\pi j_{1x}/D_{t}$. As a result, we obtain%
\begin{equation}
\frac{4\pi }{D_{t}}\left( 1+\frac{D_{t}\nu _{ie}}{\omega _{pi}^{2}}%
a_{xx}\right) j_{1x}=a_{xx}E_{1x}-a_{xy}E_{1y}-a_{xz}E_{1z}+\frac{4\pi \nu
_{ie}}{\omega _{pi}^{2}}a_{xy}j_{1y}+\frac{4\pi \nu _{ie}}{\omega _{pi}^{2}}%
a_{xz}j_{1z}.
\end{equation}%
The following notations are introduced here:%
\begin{eqnarray}
a_{xx} &=&\frac{\omega _{pi}^{2}}{\omega _{ci}^{2}}\left( \frac{\partial ^{2}%
}{\partial y^{2}}+\frac{\partial ^{2}}{\partial z^{2}}\right) \left( \frac{%
\partial }{\partial z}\right) ^{-2},a_{xy}=\frac{\omega _{pi}^{2}}{\omega
_{ci}^{2}}\frac{\partial ^{2}}{\partial x\partial y}\left( \frac{\partial }{%
\partial z}\right) ^{-2}, \\
a_{xz} &=&\frac{\omega _{pi}^{2}}{\omega _{ci}^{2}}\left[ \frac{\omega
_{ci}D_{t}}{c_{si}^{2}}\frac{\partial }{\partial y}\left( \frac{\partial }{%
\partial z}\right) ^{-2}+\frac{\partial }{\partial x}\right] \left( \frac{%
\partial }{\partial z}\right) ^{-1}.  \nonumber
\end{eqnarray}%
When deriving Equation (40), we have used expressions (21) and (26) for $%
\mathbf{F}_{i1}$ and $\mathbf{F}_{e1}$, respectively. We note that in
Equation (40) and below in this section, we have not imposed any
restrictions on the collision frequency $\nu _{ie}$ in comparison with the
ion cyclotron frequency $\omega _{ci}$.

\bigskip\ 

\subsection{Current $j_{1y}$}

Let us consider the velocity $v_{i1y}$. Using corresponding conditions, we
can write from Equation (A18) 
\begin{eqnarray}
v_{i1y} &\mathbf{=}&\mathbf{-}\frac{1}{\omega _{ci}}F_{i1x}-\frac{D_{t}}{%
\omega _{ci}^{2}}\frac{\partial ^{2}}{\partial x\partial y}\left( \frac{%
\partial }{\partial z}\right) ^{-2}F_{i1x} \\
&&+\frac{1}{D_{t}\omega _{ci}^{2}}\left[ D_{t}^{2}\left( \frac{\partial ^{2}%
}{\partial x^{2}}+\frac{\partial ^{2}}{\partial z^{2}}\right) \left( \frac{%
\partial }{\partial z}\right) ^{-2}+\omega _{bi}^{2}\right] F_{i1y} 
\nonumber \\
&&+\frac{1}{\omega _{ci}}\frac{\partial }{\partial x}\left( \frac{\partial }{%
\partial z}\right) ^{-1}F_{i1z}+\frac{1}{\omega _{ci}}\left[ \frac{D_{t}^{2}%
}{c_{si}^{2}}\frac{\partial }{\partial x}\left( \frac{\partial }{\partial z}%
\right) ^{-2}-\frac{\omega _{bi}^{2}}{g_{i}}\right] \left( \frac{\partial }{%
\partial z}\right) ^{-1}F_{i1z}  \nonumber \\
&&-\frac{1}{D_{t}\omega _{ci}^{2}}\left( D_{t}^{2}+\omega _{bi}^{2}\right) 
\frac{\partial }{\partial y}\left( \frac{\partial }{\partial z}\right)
^{-1}F_{i1z}.  \nonumber
\end{eqnarray}%
Here, we have assumed that $\partial /\partial x\gg \left(
g_{i}/c_{si}^{2}\right) \left( \partial /\partial x\right) ^{2}\left(
\partial /\partial z\right) ^{-2}$. We keep the last term in Equation (42)
because the term proportional to $\left( \omega _{bi}^{2}/\omega
_{ci}g_{i}\right) F_{i1z}$ can be canceled with the corresponding electron
term (see below). Here and below, we have introduced the notation 
\begin{equation}
\omega _{bi,e}^{2}=\frac{g_{i,e}}{c_{si,e}^{2}}\left[ g_{i,e}\left( \gamma
-1\right) +\frac{\partial c_{si,e}^{2}}{\partial x}\right] ,
\end{equation}%
which may be named the ion (electron) Brunt-V\"{a}is\"{a}l\"{a} frequency.

The velocity $v_{e1y}$ is defined by Equation (B6). For the case under
consideration, we obtain%
\begin{eqnarray}
m_{e}\omega _{ce}v_{e1y} &\mathbf{=}&\mathbf{-}F_{e1x}+m_{i}\omega _{be}^{2}%
\frac{\gamma }{\gamma D_{t}+\Omega }\frac{1}{m_{e}\omega _{ce}}F_{e1y}+\frac{%
\partial }{\partial x}\left( \frac{\partial }{\partial z}\right) ^{-1}F_{e1z}
\\
&&-m_{i}\omega _{be}^{2}\frac{\gamma }{\gamma D_{t}+\Omega }\frac{1}{%
m_{e}\omega _{ce}}\frac{\partial }{\partial y}\left( \frac{\partial }{%
\partial z}\right) ^{-1}F_{e1z}  \nonumber \\
&&-\frac{1}{c_{se}^{2}}\left[ \left( \gamma -1\right) g_{e}\frac{\gamma D_{t}%
}{\gamma D_{t}+\Omega }+\frac{\partial c_{se}^{2}}{\partial x}\right] \left( 
\frac{\partial }{\partial z}\right) ^{-1}F_{e1z}  \nonumber \\
&&+g_{e}m_{i}\frac{\partial T_{e0}}{T_{e0}\partial x}\frac{1}{B_{0}}\frac{%
\Omega }{\gamma D_{t}+\Omega }\left( \frac{\partial }{\partial z}\right)
^{-1}B_{1x}.  \nonumber
\end{eqnarray}%
If we put $\Omega =0$ in Equation (44) and take $m_{i}\rightarrow 0$ in
Equation (42), we obtain the full conformity of both equations with each
other.

Using expressions (42) and (44), we find the current $j_{1y}$. Taking into
account (21) and (26), we obtain%
\begin{eqnarray}
\frac{4\pi }{D_{t}}\left( 1+\frac{D_{t}\nu _{ie}}{\omega _{pi}^{2}}%
a_{yy}\right) j_{1y} &=&-a_{yx}E_{1x}+\left( a_{yy}+a_{yy}^{\ast }\right)
E_{1y}+\left( a_{yz}-a_{yz}^{\ast }\right) E_{1z} \\
&&+\frac{4\pi \nu _{ie}}{\omega _{pi}^{2}}a_{yx}j_{1x}-\frac{4\pi \nu _{ie}}{%
\omega _{pi}^{2}}a_{yz}j_{1z},  \nonumber
\end{eqnarray}%
where the following notations are introduced:%
\begin{eqnarray}
a_{yx} &=&\frac{\omega _{pi}^{2}}{\omega _{ci}^{2}}\frac{\partial ^{2}}{%
\partial x\partial y}\left( \frac{\partial }{\partial z}\right) ^{-2},a_{yy}=%
\frac{\omega _{pi}^{2}}{\omega _{ci}^{2}}\left[ \left( \frac{\partial ^{2}}{%
\partial x^{2}}+\frac{\partial ^{2}}{\partial z^{2}}\right) \left( \frac{%
\partial }{\partial z}\right) ^{-2}+\frac{\omega _{bi}^{2}}{D_{t}^{2}}+\frac{%
\omega _{be}^{2}}{D_{t}^{2}}\frac{\gamma D_{t}}{\gamma D_{t}+\Omega }\right]
, \\
a_{yz} &=&\frac{\omega _{pi}^{2}}{D_{t}\omega _{ci}}\left\{ \frac{D_{t}^{2}}{%
c_{si}^{2}}\frac{\partial }{\partial x}\left( \frac{\partial }{\partial z}%
\right) ^{-2}-\frac{\omega _{bi}^{2}}{g_{i}}+\frac{1}{c_{se}^{2}}\left[
\left( \gamma -1\right) g_{e}\frac{\gamma D_{t}}{\gamma D_{t}+\Omega }+\frac{%
\partial c_{se}^{2}}{\partial x}\right] \right\} \left( \frac{\partial }{%
\partial z}\right) ^{-1}  \nonumber \\
&&-\frac{\omega _{pi}^{2}}{\omega _{ci}^{2}}\left\{ 1+\frac{\omega _{bi}^{2}%
}{D_{t}^{2}}+\frac{\gamma \omega _{be}^{2}}{D_{t}\left( \gamma D_{t}+\Omega
\right) }\right\} \frac{\partial }{\partial y}\left( \frac{\partial }{%
\partial z}\right) ^{-1},  \nonumber \\
a_{yy}^{\ast } &=&\frac{1}{D_{t}^{2}}\frac{\omega _{pi}^{2}}{\omega _{ci}^{2}%
}g_{e}\frac{\partial T_{e0}^{\ast }}{T_{e0}\partial x}\frac{\Omega }{\gamma
D_{t}+\Omega },a_{yz}^{\ast }=a_{yy}^{\ast }\frac{\partial }{\partial y}%
\left( \frac{\partial }{\partial z}\right) ^{-1}.  \nonumber
\end{eqnarray}%
The terms proportional to $a_{yy,z}^{\ast }$ are connected with the
contribution of the heat flux $\sim B_{1x}$ (see Eq. [32]). This magnetic
field perturbation has been expressed via electric field perturbation
through Equation (9).

We see from the expression for $a_{yz}$ that for $\Omega =0$ and $%
T_{i0}=T_{e0}$ the last two terms in the first figured brackets are canceled.

\bigskip

\subsection{Current $j_{1z}$}

From Equation (A20), we find the simplified ion velocity $v_{i1z}$%
\begin{eqnarray}
\frac{\partial v_{i1z}}{\partial z} &\mathbf{=}&\frac{1}{\omega _{ci}}\frac{%
\partial F_{i1x}}{\partial y}+\frac{D_{t}^{2}}{\omega _{ci}c_{si}^{2}}\frac{%
\partial }{\partial y}\left( \frac{\partial }{\partial z}\right)
^{-2}F_{i1x}-\frac{D_{t}}{\omega _{ci}^{2}}\frac{\partial F_{i1x}}{\partial x%
}-\frac{1}{\omega _{ci}}\frac{\partial F_{i1y}}{\partial x}+\frac{g_{i}}{%
\omega _{ci}c_{si}^{2}}F_{i1y} \\
&&-\frac{D_{t}^{2}}{\omega _{ci}c_{si}^{2}}\frac{\partial }{\partial x}%
\left( \frac{\partial }{\partial z}\right) ^{-2}F_{i1y}-\frac{1}{D_{t}\omega
_{ci}^{2}}\left( D_{t}^{2}+\omega _{bi}^{2}\right) \frac{\partial F_{i1y}}{%
\partial y}\mathbf{-}\frac{D_{t}}{c_{si}^{2}}\left( \frac{\partial }{%
\partial z}\right) ^{-1}F_{i1z}.  \nonumber
\end{eqnarray}%
We have taken into account that $\partial /\partial x\gg \left(
g_{i}/c_{si}^{2}\right) \left[ \left( \partial /\partial x\right)
^{2}+\left( \partial /\partial y\right) ^{2}\right] \left( \partial
/\partial z\right) ^{-2}$. From Equation (B8), we have 
\begin{eqnarray}
\frac{\partial v_{e1z}}{\partial z} &=&\frac{1}{m_{e}\omega _{ce}}\frac{%
\partial F_{e1x}}{\partial y}-\frac{1}{m_{e}\omega _{ce}}\frac{\partial
F_{e1y}}{\partial x}+\frac{1}{c_{se1}^{2}}\left[ g_{e}+\frac{\Omega }{\left(
D_{t}+\Omega \right) }\frac{\partial T_{e0}}{m_{i}\partial x}\right] \frac{1%
}{m_{e}\omega _{ce}}F_{e1y} \\
&&-\frac{\gamma \omega _{be}^{2}m_{i}}{\left( \gamma D_{t}+\Omega \right) }%
\left( \frac{1}{m_{e}\omega _{ce}}\right) ^{2}\frac{\partial F_{e1y}}{%
\partial y}\mathbf{-}\frac{D_{t}}{c_{se1}^{2}m_{i}}\left( \frac{\partial }{%
\partial z}\right) ^{-1}F_{e1z}  \nonumber \\
&&-\frac{\partial T_{e0}}{c_{se1}^{2}m_{i}\partial x}\frac{1}{B_{0}}\frac{%
D_{t}\Omega }{\left( D_{t}+\Omega \right) }\left( \frac{\partial }{\partial z%
}\right) ^{-1}B_{1x},  \nonumber
\end{eqnarray}%
where 
\begin{equation}
c_{se1}^{2}=c_{se}^{2}\frac{\left( \gamma D_{te}+\Omega \right) }{\gamma
\left( D_{te}+\Omega \right) }.
\end{equation}%
If we take in Equation (47) $m_{i}\rightarrow 0$ and put $\Omega =0$ in
Equation (48), we obtain the conformity of these two equations.

Using Equations (9), (47), and (48), we find the longitudinal current $%
j_{1z} $%
\begin{eqnarray}
\frac{4\pi }{D_{t}}\left( 1-\frac{D_{t}\nu _{ie}}{\omega _{pi}^{2}}%
a_{zz}\right) j_{1z} &=&a_{zx}E_{1x}+\left( a_{zy}+a_{zy}^{\ast }\right)
E_{1y}-\left( a_{zz}+a_{zz}^{\ast }\right) E_{1z}-\frac{4\pi \nu _{ie}}{%
\omega _{pi}^{2}}a_{zx}j_{1x} \\
&&-\frac{4\pi \nu _{ie}}{\omega _{pi}^{2}}a_{zy}j_{1y}.  \nonumber
\end{eqnarray}%
The following notations are introduced here:%
\begin{eqnarray}
a_{zx} &=&\frac{\omega _{pi}^{2}}{\omega _{ci}^{2}}\left[ \frac{\omega
_{ci}D_{t}}{c_{si}^{2}}\frac{\partial }{\partial y}\left( \frac{\partial }{%
\partial z}\right) ^{-2}-\frac{\partial }{\partial x}\right] \left( \frac{%
\partial }{\partial z}\right) ^{-1},a_{zz}=\omega _{pi}^{2}\left( \frac{1}{%
c_{si}^{2}}+\frac{1}{c_{se1}^{2}}\right) \left( \frac{\partial }{\partial z}%
\right) ^{-2}, \\
a_{zy} &=&\frac{\omega _{pi}^{2}}{\omega _{ci}D_{t}}\left\{ \frac{g_{i}}{%
c_{si}^{2}}-\frac{1}{c_{se1}^{2}}\left[ g_{e}+\frac{\Omega }{\left(
D_{t}+\Omega \right) }\frac{\partial T_{e0}}{m_{i}\partial x}\right] -\frac{%
D_{t}^{2}}{c_{si}^{2}}\frac{\partial }{\partial x}\left( \frac{\partial }{%
\partial z}\right) ^{-2}\right\} \left( \frac{\partial }{\partial z}\right)
^{-1}  \nonumber \\
&&-\frac{\omega _{pi}^{2}}{\omega _{ci}^{2}D_{t}^{2}}\left( D_{t}^{2}+\omega
_{bi}^{2}+\omega _{be}^{2}\frac{\gamma D_{t}}{\gamma D_{t}+\Omega }\right) 
\frac{\partial }{\partial y}\left( \frac{\partial }{\partial z}\right) ^{-1},
\nonumber \\
a_{zy}^{\ast } &=&\frac{\omega _{pi}^{2}}{\omega _{ci}D_{t}}\frac{\partial
T_{e0}^{\ast }}{c_{se1}^{2}m_{i}\partial x}\frac{\Omega }{\left(
D_{t}+\Omega \right) }\left( \frac{\partial }{\partial z}\right)
^{-1},a_{zz}^{\ast }=a_{zy}^{\ast }\frac{\partial }{\partial y}\left( \frac{%
\partial }{\partial z}\right) ^{-1}.  \nonumber
\end{eqnarray}%
The terms proportional to $a_{zy,z}^{\ast }$ in Equation (50) are connected
with the contribution of the electron current induced by $B_{1x}$ (see Eq.
[48]).

\bigskip

\section{SIMPLIFIED\ COLLISION\ CONTRIBUTION}

\bigskip

To take into account collisions between ions and electrons, we involve some
simplifying assumptions. First of all, we assume that 
\begin{equation}
1\gg \frac{D_{t}\nu _{ie}}{\omega _{ci}^{2}}.
\end{equation}%
This condition allows us to neglect the collision contribution on the left
hand-side of Equations (40) and (45). We note that the relation between $%
\omega _{ci}$ and $\nu _{ie}$ in inequality (52) can be arbitrary because $%
D_{t}\ll \omega _{ci}$. For estimations of contribution of the different
terms, we assume here and below that $D_{t}\gtrsim \omega _{bi,e}$. We also
consider that%
\begin{equation}
\max \left\{ \frac{D_{t}^{2}}{c_{si}^{2}}\frac{\partial }{\partial y}\left( 
\frac{\partial }{\partial z}\right) ^{-2};\frac{D_{t}}{\omega _{ci}}\frac{%
\partial }{\partial x}\right\} L\gg \frac{D_{t}\nu _{ie}}{\omega _{ci}^{2}}%
\frac{\partial ^{2}}{\partial x\partial y}\left( \frac{\partial }{\partial z}%
\right) ^{-2},
\end{equation}%
where $L$ is the typical inhomogeneity scale length along the $x$-axis.
Under conditions (52) and (53), we can neglect the contribution of the
collisional terms proportional to $\nu _{ie}^{2}$. Then solving Equations
(40), (45), and (50), we can express the current $\mathbf{j}_{1}$ through $%
\mathbf{E}_{1}$%
\begin{eqnarray}
\frac{4\pi }{D_{t}}j_{1x} &=&\varepsilon _{xx}E_{1x}-\varepsilon
_{xy}E_{1y}-\varepsilon _{xz}E_{1z}, \\
\frac{4\pi }{D_{t}}j_{1y} &=&-\varepsilon _{yx}E_{1x}+\varepsilon
_{yy}E_{1y}+\varepsilon _{yz}E_{1z},  \nonumber \\
\frac{4\pi }{D_{t}}j_{1z} &=&\varepsilon _{zx}E_{1x}+\varepsilon
_{zy}E_{1y}-\varepsilon _{zz}E_{1z}.  \nonumber
\end{eqnarray}%
The following notations are introduced here:%
\begin{eqnarray}
\varepsilon _{xx} &=&a_{xx},\varepsilon _{xy}=a_{xy}-\frac{D_{t}\nu _{ie}}{%
\omega _{pi}^{2}}\frac{a_{xz}\left( a_{zy}+a_{zy}^{\ast }\right) }{\left(
1-id_{z}\right) }, \\
\varepsilon _{xz} &=&a_{xz}-\frac{D_{t}\nu _{ie}}{\omega _{pi}^{2}}\left[
a_{xy}\left( a_{yz}-a_{yz}^{\ast }\right) -\frac{a_{xz}\left(
a_{zz}+a_{zz}^{\ast }\right) }{\left( 1-id_{z}\right) }\right] ,  \nonumber
\\
\varepsilon _{yx} &=&a_{yx}+\frac{D_{t}\nu _{ie}}{\omega _{pi}^{2}}\frac{%
a_{yz}a_{zx}}{\left( 1-id_{z}\right) },\varepsilon _{yy}=a_{yy}+a_{yy}^{\ast
}-\frac{D_{t}\nu _{ie}}{\omega _{pi}^{2}}\frac{a_{yz}\left(
a_{zy}+a_{zy}^{\ast }\right) }{\left( 1-id_{z}\right) },  \nonumber \\
\varepsilon _{yz} &=&a_{yz}-a_{yz}^{\ast }+\frac{D_{t}\nu _{ie}}{\omega
_{pi}^{2}}\frac{a_{yz}\left( a_{zz}+a_{zz}^{\ast }\right) }{\left(
1-id_{z}\right) },  \nonumber \\
\varepsilon _{zx} &=&\frac{a_{zx}}{1-id_{z}},\varepsilon _{zy}=\frac{%
a_{zy}+a_{zy}^{\ast }}{1-id_{z}},\varepsilon _{zz}=\frac{a_{zz}+a_{zz}^{\ast
}}{1-id_{z}},  \nonumber
\end{eqnarray}%
where 
\begin{equation}
id_{z}=a_{zz}\frac{D_{t}\nu _{ie}^{0}}{\omega _{pi}^{2}}.
\end{equation}%
We keep in $\varepsilon _{ij}$ ($i,j=x,y,z$) some terms, which will turn out
to be small, because we do not know \textit{a priori }solution of the
dispersion relation.

\bigskip

\section{DISPERSION\ RELATION}

\bigskip

To derive the dispersion relation, we apply the Fourier transformation to
the electromagnetic field and current, assuming that $\mathbf{E}_{1}\mathbf{%
\sim E}_{1k}\exp \left( i\mathbf{kr-}i\omega t\right) $, where $k=\left\{ 
\mathbf{k},\omega \right\} $. We take into account all three components of
the wave vector, $\mathbf{k}=\left( k_{x},k_{y},k_{z}\right) $. Then using
Equations (9), (10), and (54), we obtain%
\begin{equation}
\mathbf{\hat{A}}_{k}\mathbf{E}_{1k}=\mathbf{0,}
\end{equation}%
where $\mathbf{E}_{1k}=\left( E_{1xk},E_{1yk},E_{1zk}\right) $ and the
matrix $\mathbf{\hat{A}}_{k}$ has the form%
\[
\mathbf{\hat{A}}_{k}=\left\vert 
\begin{array}{ccc}
n_{y}^{2}+n_{z}^{2}-\varepsilon _{xx}, & -n_{x}n_{y}+\varepsilon _{xy}, & 
-n_{x}n_{z}+\varepsilon _{xz} \\ 
-n_{x}n_{y}+\varepsilon _{yx}, & n_{x}^{2}+n_{z}^{2}-\varepsilon _{yy}, & 
-n_{y}n_{z}-\varepsilon _{yz} \\ 
-n_{x}n_{z}-\varepsilon _{zx}, & -n_{y}n_{z}-\varepsilon _{zy}, & 
n_{x}^{2}+n_{y}^{2}+\varepsilon _{zz}%
\end{array}%
\right\vert . 
\]%
Here $\mathbf{n}=\mathbf{k}c/\omega $. We keep for values $\varepsilon _{ij}$
the same notations as above. The dispersion relation is the determinant of
the matrix $\mathbf{\hat{A}}_{k}$ equal to zero%
\begin{eqnarray}
0 &=&\left( n_{y}^{2}+n_{z}^{2}-\varepsilon _{xx}\right) \left[ \left(
n_{x}^{2}+n_{z}^{2}-\varepsilon _{yy}\right) \varepsilon _{zz}-\varepsilon
_{zy}\varepsilon _{yz}\right] \\
&&-\left( n_{x}n_{y}-\varepsilon _{yx}\right) \left[ \left(
n_{x}n_{y}-\varepsilon _{xy}\right) \varepsilon _{zz}-\varepsilon
_{xz}\varepsilon _{zy}+\varepsilon _{zx}\varepsilon _{yz}\right] .  \nonumber
\end{eqnarray}%
When obtaining Equation (58), we have taken into account that according to
Equations (41), (46), (51), and (55) inequalities $\varepsilon _{zz}\gg
\varepsilon _{xx},\varepsilon _{xy},\varepsilon _{xz},\varepsilon
_{yy},\varepsilon _{yz},\varepsilon _{zy}$ and $\varepsilon _{xx}\varepsilon
_{zz}\gg \varepsilon _{xz}\varepsilon _{zx}$ are satisfied. Expression $%
\varepsilon _{xz}\varepsilon _{zy}-\varepsilon _{zx}\varepsilon _{yz}$ can
be omitted under the condition $1\gg 1/\left( 1-id_{z}\right) kL$, when $%
\Omega \gtrsim \omega $ or $T_{i}\neq T_{e}$ at $\Omega \ll \omega $. Here,
the value $k$ is equal to $k=\left\vert \mathbf{k}\right\vert $ and $\Omega
=\left( \gamma -1\right) \chi _{e0}k_{z}^{2}/n_{e0}$ (see Eq. [35]).

To further simplify the dispersion relation (58), we neglect the collision
contribution in $\varepsilon _{xy}$ and $\varepsilon _{yx}$. The
corresponding condition can be written in the form%
\begin{equation}
1\gg \frac{1}{\left( 1-id_{z}\right) }\left( id_{z}+\frac{\nu _{ie}}{\omega
_{ci}}\right) \left( \frac{1}{kL}+\frac{\omega ^{2}}{k_{z}^{2}c_{si}^{2}}%
\right) .
\end{equation}%
We remind that the ratio $\nu _{ie}/\omega _{ci}$ in this inequality is
arbitrary. Then Equation (58) takes the form%
\begin{equation}
\omega ^{2}=k_{z}^{2}c_{A}^{2}+\frac{\omega _{ci}^{2}}{\omega _{pi}^{2}}%
D_{t}^{2}\left( \varepsilon _{yy1}+\frac{\varepsilon _{zy}\varepsilon _{yz}}{%
\varepsilon _{zz}}\right) \frac{\left( k_{y}^{2}+k_{z}^{2}\right) }{\left(
k_{x}^{2}+k_{y}^{2}+k_{z}^{2}\right) },
\end{equation}%
where $c_{A}=\left( B_{0}/4\pi m_{i}n_{i0}\right) ^{1/2}$ is the ion Alfv%
\'{e}n velocity. The following notation is introduced here: 
\begin{equation}
\varepsilon _{yy1}=\frac{\omega _{pi}^{2}}{\omega _{ci}^{2}}\left[ \frac{%
\omega _{bi}^{2}}{D_{t}^{2}}+\frac{\omega _{be}^{2}}{D_{t}^{2}}\frac{\gamma
D_{t}}{\gamma D_{t}+\Omega }\right] +a_{yy}^{\ast }-\frac{D_{t}\nu _{ie}}{%
\omega _{pi}^{2}}\frac{a_{yz}\left( a_{zy}+a_{zy}^{\ast }\right) }{\left(
1-id_{z}\right) }.
\end{equation}%
For convenience, we retain the symbol $D_{t}=-i\omega $ on the right
hand-side of Equation (60). Below, we consider the dispersion relation in
the collisionless as well as in the collisional cases.

\bigskip

\subsection{Collisionless case}

In this case, we assume that 
\begin{equation}
d_{z}\ll 1,
\end{equation}%
where $d_{z}$ is defined by expression (56). Under condition (62), the
collisional term in $\varepsilon _{yy1}$ is unimportant. Then the dispersion
relation is given by 
\begin{equation}
\omega ^{2}=k_{z}^{2}c_{A}^{2}+Wc_{s}^{2}\frac{\left(
k_{y}^{2}+k_{z}^{2}\right) }{\left( k_{x}^{2}+k_{y}^{2}+k_{z}^{2}\right) },
\end{equation}%
where $c_{s}^{2}=c_{si}^{2}c_{se1}^{2}/\left( c_{si}^{2}+c_{se1}^{2}\right) $
and 
\begin{eqnarray}
W &=&\frac{1}{c_{s}^{2}}\left( \omega _{bi}^{2}+\omega _{be}^{2}\frac{\gamma
D_{t}}{\gamma D_{t}+\Omega }+g_{e}\frac{\partial T_{e0}^{\ast }}{%
T_{e0}\partial x}\frac{\Omega }{\gamma D_{t}+\Omega }\right) \\
&&+\left\{ -\frac{\omega _{bi}^{2}}{g_{i}}+\frac{1}{c_{se}^{2}}\left[ \left(
\gamma -1\right) g_{e}\frac{\gamma D_{t}}{\gamma D_{t}+\Omega }+\frac{%
\partial c_{se}^{2}}{\partial x}\right] \right\} \left( \frac{g_{i}}{%
c_{si}^{2}}-\frac{g_{e}}{c_{se1}^{2}}\right) .  \nonumber
\end{eqnarray}%
The condition 
\[
1\gg \frac{\nu _{ie}}{\omega _{ci}}\frac{k_{y}}{Lk_{z}^{2}}\frac{\Omega }{%
\left( D_{t}+\Omega \right) } 
\]%
has been supposed.

\bigskip

\paragraph{The case $\Omega \gg D_{t}$}

In this limit, the value $W$ is the following:%
\[
W=\frac{1}{c_{s}^{2}}\left( \omega _{bi}^{2}+g_{e}\frac{\partial
T_{e0}^{\ast }}{T_{e0}\partial x}\right) +\left( -\frac{\omega _{bi}^{2}}{%
g_{i}}+\frac{1}{c_{se}^{2}}\frac{\partial c_{se}^{2}}{\partial x}\right)
\left( \frac{g_{i}}{c_{si}^{2}}-\frac{\gamma g_{e}}{c_{se}^{2}}\right) . 
\]%
This expression can be rewritten in the form 
\begin{equation}
W=\gamma \frac{\left( g_{i}+g_{e}\right) }{c_{si}^{2}c_{se}^{2}}\left[
\left( \gamma -1\right) g_{i}+\frac{1}{m_{i}}\left( \gamma \frac{\partial
T_{i0}}{\partial x}+\frac{\partial T_{e0}}{\partial x}\right) \right] ,
\end{equation}%
where we have used expression (43) for $\omega _{bi,e}^{2}$. We see that
instability, $W<0$, is possible, if the temperature increases along the
gravity (we assume that $g_{i,e}$ has the same sign as $g$).

\bigskip

\paragraph{The case $\Omega \ll D_{t}$}

In the case $\Omega \ll D_{i}$, the thermal conductivity is absent. The
value $W$ is given by%
\[
W=\frac{1}{c_{s}^{2}}\left( \omega _{bi}^{2}+\omega _{be}^{2}\right) -\left( 
\frac{\omega _{bi}^{2}}{g_{i}}-\frac{\omega _{be}^{2}}{g_{e}}\right) \left( 
\frac{g_{i}}{c_{si}^{2}}-\frac{g_{e}}{c_{se}^{2}}\right) 
\]%
or%
\begin{equation}
W=\frac{\left( g_{i}+g_{e}\right) }{c_{si}^{2}c_{se}^{2}}\left[ \left(
\gamma -1\right) \left( g_{i}+g_{e}\right) +\frac{\gamma }{m_{i}}\left( 
\frac{\partial T_{i0}}{\partial x}+\frac{\partial T_{e0}}{\partial x}\right) %
\right] .
\end{equation}

Comparing Equations (65) and (66), we see that the thermal conductivity is
not of fundamental importance.

\bigskip

\subsection{Collisional case}

We now assume that 
\begin{equation}
id_{z}\gg 1.
\end{equation}%
In this case, we can neglect the term $\varepsilon _{zy}\varepsilon _{yz}/$ $%
\varepsilon _{zz}$ in Equation (60). However, the collisional term in
expression (61) gives the same contribution as other terms. As a result, we
obtain again Equation (63) with $W$ defined by Equation (64). Thus, the
dispersion relation is the same for both the collisionless and collisional
cases. We note that this result has also been obtained for the case in which
gravity is parallel to the magnetic field (Nekrasov \& Shadmehri 2010).

\bigskip

\subsection{Polarization of the electric field perturbation}

The dispersion relation (58) without $\varepsilon _{zx}\varepsilon _{yz}$ $%
-\varepsilon _{xz}\varepsilon _{zy}$ can be obtained, if we neglect some
terms in the matrix $\mathbf{\hat{A}}_{k}$%
\[
\mathbf{\hat{A}}_{k}=\left\vert 
\begin{array}{ccc}
n_{y}^{2}+n_{z}^{2}-\varepsilon _{xx}, & -n_{x}n_{y}+\varepsilon _{xy}, & 0
\\ 
-n_{x}n_{y}+\varepsilon _{yx}, & n_{x}^{2}+n_{z}^{2}-\varepsilon _{yy}, & 
-\varepsilon _{yz} \\ 
0 & -\varepsilon _{zy}, & \varepsilon _{zz}%
\end{array}%
\right\vert . 
\]%
Then Equation (57) takes the form 
\begin{eqnarray}
\left( n_{y}^{2}+n_{z}^{2}-\varepsilon _{xx}\right) E_{1xk}+\left(
-n_{x}n_{y}+\varepsilon _{xy}\right) E_{1yk} &=&0, \\
\left( -n_{x}n_{y}+\varepsilon _{yx}\right) E_{1xk}+\left(
n_{x}^{2}+n_{z}^{2}-\varepsilon _{yy}\right) E_{1yk}-\varepsilon
_{yz}E_{1zk} &=&0,  \nonumber \\
-\varepsilon _{zy}E_{1yk}+\varepsilon _{zz}E_{1zk} &=&0.  \nonumber
\end{eqnarray}%
If we put $n_{x}=0$, then the magnetosonic wave $\omega
^{2}=k_{z}^{2}c_{A}^{2}$ is split, and Equation (60) describes the Alfv\'{e}%
n type wave. When $n_{y}=0$, the Alfv\'{e}n wave $\omega
^{2}=k_{z}^{2}c_{A}^{2}$ is split, and Equation (60) describes the
magnetosonic type wave. In the case $n_{x}=n_{y}=0$, both waves are of the
Alfv\'{e}n type. In a general case, we have $\omega ^{2}\neq
k_{z}^{2}c_{A}^{2}$ and $E_{1xk}=$ $\left[ k_{x}k_{y}/\left(
k_{y}^{2}+k_{z}^{2}\right) \right] E_{1yk}$.

We see from the system of equations (68) that $E_{1zk}=\left( \varepsilon
_{zy}/\varepsilon _{zz}\right) E_{1yk}\ll E_{1yk}$. However, the
contribution of the longitudinal electric field $E_{1zk}$ must be taken into
account in the collisionless as well as in the collisional cases (see Eqs.
[60] and [61]).

\bigskip

\section{DISCUSSION}

\bigskip

The dispersion relation (60) emphasizes the important role of the perturbed
current and electric field along the background magnetic field. These
perturbations produce the terms connected with $\varepsilon _{yz}$, $%
\varepsilon _{zy}$, and $\varepsilon _{zz}$ in the dispersion. Taking into
account these terms allows us to derive the correct dispersion relation
which has the form (63). The value $W$, defined by expression (64), is
available for both cases when the electron thermal conductivity is present
or absent. Expressions (65) and (66) show that the thermal conductivity is
not of fundamental importance for the buoyancy instability, if the latter
can be excited. For the instability, the temperature gradient of ions and
electrons must have the sign opposite to that of $g_{i,e}$. Under assumption
that $g_{i}$ and $g_{e}$ have the same sign as $g$, the temperature must
increase along the gravity for the instability to be excited.

The dispersion relation (63) takes into account collisions. Except for the
anisotropic thermal conductivity adopted in this paper, the relation between 
$\omega _{ci}$ and $\nu _{ie}$ can be arbitrary in the framework of
conditions (52), (53), and (59). Under these conditions, the dispersion
relation for the collisionless (62) and collisional (67) cases is the same.
This result has also been obtained for the case in which the background
magnetic field and gravity are parallel to each other (Nekrasov \& Shadmehri
2010).

From Equations (23) and (24), we can find the ion number density and
pressure perturbations. Using Equations (38), (42), and (47), we can
calculate the value $\mathbf{\nabla }\cdot \mathbf{v}_{i1}$. Keeping the
main terms, we have

\begin{equation}
\mathbf{\nabla }\cdot \mathbf{v}_{i1}\simeq \frac{g_{i}}{\omega
_{ci}c_{si}^{2}}F_{i1y}-\frac{D_{t}}{c_{si}^{2}}\left( \frac{\partial }{%
\partial z}\right) ^{-1}F_{i1z},
\end{equation}%
Substituting $v_{i1x}$ and $\mathbf{\nabla }\cdot \mathbf{v}_{i1}$ in
Equation (23), we obtain an estimation 
\begin{equation}
-D_{t}\frac{n_{i1}}{n_{i0}}\simeq \frac{1}{\omega _{ci}}\left( \frac{1}{%
n_{i0}}\frac{\partial n_{i0}}{\partial x}+\frac{g_{i}}{c_{si}^{2}}\right)
F_{i1y}-\frac{D_{t}}{c_{si}^{2}}\left( \frac{\partial }{\partial z}\right)
^{-1}F_{i1z}.
\end{equation}%
We see from Equations (69) and (70) that $\mathbf{\nabla }\cdot \mathbf{v}%
_{i1}\sim D_{t}n_{i1}/n_{i0}$. From the system of equations (68), it is
followed that $E_{1zk}=\left( \varepsilon _{zy}/\varepsilon _{zz}\right)
E_{1yk}$. Thus, both terms on the right hand-side of equation (70) are of
the same order. An estimation for the ion pressure perturbation is given by%
\[
D_{t}\frac{p_{i1}}{\gamma p_{i0}}\simeq \frac{D_{t}}{c_{si}^{2}}\left( \frac{%
\partial }{\partial z}\right) ^{-1}F_{i1z}. 
\]%
We see that $n_{i1}/n_{i0}\sim p_{i1}/p_{i0}$ due to the longitudinal
electric field perturbation. The analogous conclusion can also be made for
electrons. This result contradicts an assumption that $n_{1}/n_{0}\gg
p_{1}/p_{0}$ which one uses in the MHD analysis of buoyancy instabilities.
We note that the ideal MHD does not involve the field $E_{z}$.

Let us discuss the relevance of conditions used in this paper to real
astrophysical systems. As an example, we will consider an intracluster
medium. However, our conditions have a more general applicability.
Observations show that all astrophysical objects in cosmic space have
magnetic fields of $\mu $G strength in galaxy clusters and molecular clouds
(e.g., Carilli \& Taylor 2002; Wardle \& Ng 1999) to G and more in acretion
disks (e.g., Desch 2004; Donati et al. 2005).\textbf{\ }For such magnetic
fields, the ion Larmor radius $\rho _{i}$\ is considerably smaller than the
typical inhomogeneity length $L$ in these objects, i.e. $\rho _{i}\ll L$.
For example, if we take for ICM $B_{0}\sim 1$ $\mu $G and $T_{i}\sim 1$ keV,
we obtain $\rho _{i}\sim 2\times 10^{4}$ km. The magnitude of $L$ is $\sim $%
tens of kpc. In ICM, the collision frequency $\nu _{ie}$ is much less than
the ion cyclotron frequency $\omega _{ci}$, $\nu _{ie}\ll \omega _{ci}$. In
dense molecular clouds and accretion disks, the ions can be unmagnetized, $%
\nu _{ie}\gg \omega _{ci}$ (Wardle \& Ng 1999). However, many conditions of
our consideration (see below) can also be satisfied in the last case. For
example,\textbf{\ }this relates to inequality (17) justifying solutions (13)
and (14) in the equilibrium state. The electromagnetic buoyancy
perturbations have a dynamical frequency $\omega $ and wavelength $\lambda $
much less than the ion cyclotron and sound frequencies and inhomogeneity
scale length, respectively. Thus, the conditions (52) and (59) are satisfied
for a weakly collisional plasmas as ICM and can be satisfied for a strong
collisional plasma when $\nu _{ie}\gtrsim \omega _{ci}$. Accounting for the
fact that $\left\vert \omega \right\vert \sim g/c_{s}$\ (indices $i$\ and $e$%
\ are omitted), we can treat the medium as a stationary one (see Eq. [18]),
if $1\gg \left( \nu _{ie}/\omega _{ci}\right) \left( \rho _{i}/L\right) $
that is justified for astrophysical objects. The condition (20) is satisfied
for the last inequality (which also relates to Equation (28)) and under $%
\rho _{i}\ll L$ in the case $v_{i0y}\neq \mathrm{const}$. The condition (53)
can be written in the form $1\gg \left( \nu _{ie}/\omega _{ci}\right) \left(
\lambda /L;\rho _{i}/\lambda \right) $, where $\lambda \ll L$ and $\rho
_{i}\ll \lambda $. The first condition (37) is the following: $1\gg \left(
\rho _{i}/\lambda \right) \left( 1;\nu _{ie}/\omega _{ci}\right) $. It is
also true for the transverse component of the second condition (37).
However, taking into account that $E_{1zk}=\left( \varepsilon
_{zy}/\varepsilon _{zz}\right) E_{1yk}$ (see Sec. 8.3), both sides of the $z$%
-component of the second inequality (37) can be of the same order, if $%
\mathbf{v}_{i,e0}\neq \mathbf{0}$. Owing to indefiniteness of the background
velocities also containing the electric field, we here do not consider their
effect. Nevertheless, the streaming instabilities can also take place.

In our analysis, we consider for generality that ions and electrons have the
different temperatures. However, in Eqs. (3), (6), and (7), the terms
describing the energy exchange between species due to their collisions have
not been taken into account. This is possible, if the dynamical time scale
is smaller than the time scale of smoothing of the ion and electron
temperatures, i.e. $\nu _{ie}\ll \omega $. In the opposite case, $\nu
_{ie}\gg \omega $, the perturbed temperatures of electrons and ions are
almost equal each other. Equations (6) and (7) for electrons will keep their
form because $\mathbf{v}_{e1}\approx \mathbf{v}_{i1}$. In the case $%
T_{e0}\approx T_{i0}$, these equations will stay the same with the heat flux
two times less than the former one. Equation for the ion temperature will
not be needed.

\bigskip

\section{IMPLICATIONS OF THE OBTAINED RESULTS FOR GALAXY CLUSTERS}

\bigskip

According to a simplified point of view, the core of cluster must be cooled.
In fact, observations show that in most clusters the cooling time-scale near
the center of cluster, $10^{8}$ to $10^{9}$ yr, is much shorter than a
cluster's age $10^{10}$ yr (e.g., Fabian 1994; Peres et al. 1998; Allen
2000). But such high rates of cooling has been definitely ruled out by the
X-ray observations showing that the cluster core is sufficiently hot (e.g.,
Allen 2000). Thus, some heating mechanisms are in operation, though we have
little knowledge about them. Different heating mechanisms from AGN feedback
to cosmic rays and turbulence are proposed to resolve the "cooling flow
problem" (e.g., Eilek 2004; Binney \& Tabor 1995; Loewenstein et al. 1991;
Reynolds 2002). Exactly for this reason, a study of buoyancy instabilities
in ICM has a purpose to find a solution of this longstanding problem in
clusters\textbf{\ }of galaxies. The plasma in ICM is turbulent (e.g.,
Loewenstein \& Fabian 1990; Cattaneo \& Teyssier 2007). Buoyancy
instabilities could be one of possible sources of turbulence resulting in
the emergence\textbf{\ }of heat fluxes along the magnetic field in the
direction of core. Thus, the latter could be heated. The total nonlinear
picture of this process which can include reorientation of the magnetic
field (Parrish \& Stone 2007) is very complex for the\textbf{\ }analytical
consideration and can only be investigated numerically.

In the framework of the linear ideal MHD, new buoyancy instabilities have
been found, when the thermal conduction is the dominant mode of heat
transport (Balbus 2000, 2001; Quataert 2008). In the present paper, in the
framework of the multicomponent $\mathbf{E}$-approach, we have found
generalized growth rates (for the same geometry as considered by Balbus
2000) for both the negligible and dominant thermal conduction. In both
cases, the growth rates have the same order of magnitude. However, our
modified conditions of instability are different from the MHD case\textbf{\ }%
because of the multifluid nature of the system (see Section 9). We have
shown that conditions for the buoyancy instability have the form which%
\textbf{\ }is analogous\textbf{\ }to the Schwarzschild criterion
(Schwarzschild 1958). Thus, the ICM plasma can be buoyantly unstable\textbf{%
\ }for both large and small thermal conductivity. This increases the range
of wavelengths of the unstable perturbations because the thermal flux is
proportional to the wave number. Our linear analysis shows that the
multifluid nature of plasma in the ICM and other astrophysical objects can
not be neglected when one investigates heat flows as a result of buoyancy
instabilities. We also note that buoyancy instabilities are of the
electromagnetic nature. Therefore, they can contribute to the magnetic field
activity in astrophysical objects.

\bigskip

\section{CONCLUSION}

\bigskip

In this paper, we have investigated buoyancy instabilities in the magnetized
electron-ion astrophysical plasmas in which the background magnetic field
and gravity are perpendicular to each other. We have applied the
multicomponent $\mathbf{E}$-approach in which the dynamical equations for
ions and electrons are solved separately via the electric field
perturbations. The perturbed current and Faraday's and Ampere's laws have
been used to derive the dispersion relation. We have included collisions
between electrons and ions. Except for the anisotropy of the electron heat
flux adopted in this paper, in other respects, the relation between the
ion-electron collision frequency and ion cyclotron frequency can be
arbitrary in the framework of approximations, which have been made. The
important role of the longitudinal electric field perturbations, which are
not captured by the MHD equations, has been shown. The obtained growth rates
for cases of strong and weak electron thermal conductivity show that an
instability is possible when the temperature gradients of ions and electrons
are directed along the gravity. We have shown that the relative
perturbations of number density and pressure are of the same order as a
result of action of the longitudinal electric field perturbation.

Results obtained in this paper are applicable to the magnetized collisional
stratified objects and can be useful for a search of sources of turbulent
transport of energy and matter in the ICM and other astrophysical objects.

\bigskip

\section{Acknowledgments}

The authors gratefully acknowledge the anonymous referee whose insightful
and constructive comments and suggestions help to improve this paper. MS is
happy to acknowledge the hospitality of the staff and Axel Brandenburg at
NORDITA where parts of this work were done during a research visitor program.

\bigskip

\section{REFERENCES}

Allen, S. W. 2000, MNRAS, 315, 269

Balbus, S. A. 2000, ApJ, 534, 420

Balbus, S. A. 2001, ApJ, 562, 909

Balbus, S. A., \& Hawley, J. F. 1991, ApJ, 376, 214

Binney, J., \& Tabor, G. 1995, MNRAS, 276, 663

Braginskii, S. I. 1965, Rev. Plasma Phys., 1, 205

Califano, F., Hellinger, P., Kuznetsov, E., Passot, T., Sulem, P. L., \& Tr%
\'{a}vn\'{\i}\v{c}ek P.M. 2008, J. Geophys. Res., 113, A08219

Carilli, C. L., \& Taylor, G. B. 2002, ARA\&A, 40, 319

Cattaneo, A., \& Teyssier, R. 2007, MNRAS, 376, 1547

Chandran, B. D., \& Dennis, T. J. 2006, ApJ, 642, 140

Chang, P., \& Quataert, E. 2010, MNRAS, 403, 246

Desch, S. J. 2004, ApJ, 608, 509 \ 

Donati, J.-F., Paletou, F., Bouvier, J., \& Ferreira, J. 2005, Nature, 438,
466

Eilek, J. 2004, in \textit{The Riddle of Cooling Flows and Clusters of
Galaxies,} ed. Reiprich, T., Kempner, J., \& Soker, N., E13,
http://www.astro.virginia.edu/coolflow/proc.php

Fabian, A. C. 1994, ARA \& A, 32, 277

Fabian, A. C., Sanders, J. S., Taylor, G. B., Allen, S. W., Crawford, C. S.,
Johnstone, R. M.,\&\ Iwasawa, K. 2006, MNRAS, 366, 417

Gossard, E.E., \& Hooke, W.H. 1975, Waves in the Atmosphere (Amsterdam:
Elsevier Scientific Publishing Company)

Kennel, C. F., \& Sagdeev, R. Z. 1967 a, I, J. Geophys.Res., 72, 3303

Kennel, C. F., \& Sagdeev, R. Z. 1967 b, II, J. Geophys.Res., 72, 3327

Loewenstein, M., \& Fabian, A. 1990, MNRAS, 242, 120

Loewenstein, M., Zweibel, E., \& Begelman, M. 1991, ApJ, 377, 392

Narayan, R., Igumenshchev, I. V., \& Abramowicz, M. A. 2000, ApJ, 539, 798

Narayan, R., Quataert, E., Igumenshchev, I. V., \& Abramowicz, M. A. 2002,
ApJ, 577, 295

Nekrasov, A. K. 2008, Phys. Plasmas, 15\textbf{,} 032907

Nekrasov, A. K. 2009 a, Phys.Plasmas, 16, 032902

Nekrasov, A. K. 2009 b, ApJ, 695, 46

Nekrasov, A. K. 2009 c, ApJ, 704, 80

Nekrasov, A. K. 2009 d, MNRAS, 400, 1574

Nekrasov, A. K., \& Shadmehri, M. 2010, arXiv:1003.0204

Nishikawa, K., \& Wakatani, M. 1990, Plasma Physics, (Berlin Heidelberg:
Springer-Verlag)

Pedlosky, J. 1982, Geophysical Fluid Dynamics, (New York: Springer-Verlag)

Peres, C. B., et al. 1998, MNRAS, 298, 416

Quataert, E. 2008, ApJ, 673, 758

Rasera, Y., \& Chandran, B. 2008, ApJ, 685, 105

Ren, H., Wu, Z., Cao, J., \& Chu, P. K. 2009, Phys. Plasmas, 16, 102109

Reynolds, C. S. 2002, in ASP Conf. Proc., 250, \textit{Particles and Fields
in Radio Galaxies}, ed. R. A. Laing, \& K. M. Blundell (San Fransisco: ASP),
449

Sanders, J. S.,\ Fabian, A. C., Frank, K. A., Peterson, J. R., \& Russell,
H. R. 2010, MNRAS, 402, 127

Sarazin, C. L. 1988, X-Ray Emission from Clusters of Galaxies (Cambridge:

Cambridge Univ. Press)

Schekochihin, A. A., Cowley, S.C., Kulsrud, R.M., Rosin, M.S., \& Heinemann,
T. 2008, Phys. Rev. Lett., 100, 081301

Schwarzschild, M. 1958, Structure and Evolution of the Stars (New York:
Dover)

Sharma, P., Chandran, B. D. G., Quataert, E., \& Parrish, I. J. 2009, ApJ,
699, 348

Southwood, D. J., \& Kivelson, M. G. 1993, J. Geophys. Res., 98, 9181

Vedenov, A. A., \& Sagdeev, R. Z. 1958, Sov. Phys. --- Dokl., 3, 278

Vedenov, A. A., \& Sagdeev, R. Z. 1959, \textit{in} Plasma Physics and
Problem of Controlled Thermonuclear Reactions, vol. III, (English Edition),
edited by M. A. Leontovich, pp. 332-339, (New York: Pergamon)

Vranje\v{s}, J., Tanaka, M. Y., Kono, M., \& Poedts, S. 2003, Phys. Rev. E,
67, 026410

Wardle, M., \& Ng, C. 1999, MNRAS, 303, 239

\bigskip
\begin{appendix}

\begin{center}
\large{APPENDIX A}
\end{center}
\bigskip

\section{SOLUTION\ OF\ EQUATION (22)}

The components of Equation (22) are the following:%
\begin{equation}
D_{ti}v_{i1x}\mathbf{=}-\frac{1}{m_{i}n_{i0}}\frac{\partial p_{i1}}{\partial
x}-g_{i}\frac{n_{i1}}{n_{i0}}+G_{i1x}+\omega _{ci}v_{i1y},  
\end{equation}%
\begin{equation}
D_{ti}v_{i1y}\mathbf{=}-\frac{1}{m_{i}n_{i0}}\frac{\partial p_{i1}}{\partial
y}+G_{i1y}-C_{i1y}-\omega _{ci}v_{i1x},  
\end{equation}%
\begin{equation}
D_{ti}v_{i1z}\mathbf{=}-\frac{1}{m_{i}n_{i0}}\frac{\partial p_{i1}}{\partial
z}+G_{i1z}.  
\end{equation}%
Let us apply the operator $D_{ti}$ to Equation (A1). We do not differentiate
the value $1/n_{i0}$ and change $D_{ti}\partial p_{i1}/\partial x$ by $%
\partial D_{ti}p_{i1}/\partial x$ according to condition (20). Then we use
Equations (A2), (23) and (24). As a result, we obtain the following equation
connecting $v_{i1x}$ and $\mathbf{\nabla \cdot v}_{i1}$:%
\begin{eqnarray}
\left[ D_{ti}\left( D_{ti}^{2}+\omega _{ci}^{2}\right) +g_{i}\left( \omega
_{ci}\frac{\partial }{\partial y}+D_{ti}\frac{\partial }{\partial x}\right) %
\right] v_{i1x} &\mathbf{=}&D_{ti}\left[ D_{ti}G_{i1x}+\omega _{ci}\left(
G_{i1y}-C_{i1y}\right) \right]  \\
&&+\left\{ D_{ti}\left[ g_{i}\left( 1-\gamma \right) +c_{si}^{2}\frac{%
\partial }{\partial x}\right] +\omega _{ci}c_{si}^{2}\frac{\partial }{%
\partial y}\right\} \mathbf{\nabla }\cdot \mathbf{v}_{i1},  \nonumber
\end{eqnarray}%
where $c_{si}=\left( \gamma p_{i0}/m_{i}n_{i0}\right) ^{1/2}$ is the ion
sound velocity. To obtain the second equation expressing $\mathbf{\nabla
\cdot v}_{i1}$ through $v_{i1x}$, we apply the operator $D_{ti}\partial
/\partial y$ to Equation (A2), $D_{ti}\partial /\partial z$ to Equation (A3)
and add the resulting equations. Then using Equation (24), we find 
\begin{eqnarray}
\left[ D_{ti}^{2}-c_{si}^{2}\left( \frac{\partial ^{2}}{\partial y^{2}}+%
\frac{\partial ^{2}}{\partial z^{2}}\right) \right] \mathbf{\nabla }\cdot 
\mathbf{v}_{i1} &\mathbf{=}&D_{ti}\left[ \frac{\partial }{\partial z}G_{i1z}+%
\frac{\partial }{\partial y}\left( G_{i1y}-C_{i1y}\right) \right] 
\\
&&+\left[ D_{ti}\left( D_{ti}\frac{\partial }{\partial x}-\omega _{ci}\frac{%
\partial }{\partial y}\right) \mathbf{-}g_{i}\left( \frac{\partial ^{2}}{%
\partial y^{2}}+\frac{\partial ^{2}}{\partial z^{2}}\right) \right] v_{i1x}.
\nonumber
\end{eqnarray}

We introduce the following notations: \ \ \ \ \ \ 
\begin{eqnarray}
K_{1} &=&\left[ D_{ti}\left( D_{ti}^{2}+\omega _{ci}^{2}\right) +g_{i}\left(
\omega _{ci}\frac{\partial }{\partial y}+D_{ti}\frac{\partial }{\partial x}%
\right) \right] , \\
K_{2} &=&\left\{ D_{ti}\left[ g_{i}\left( 1-\gamma \right) +c_{si}^{2}\frac{%
\partial }{\partial x}\right] +\omega _{ci}c_{si}^{2}\frac{\partial }{%
\partial y}\right\} ,  \nonumber \\
M_{i1x} &=&D_{ti}\left[ D_{ti}G_{i1x}+\omega _{ci}\left(
G_{i1y}-C_{i1y}\right) \right] .  \nonumber
\end{eqnarray}%
Then Equation (A4) takes the form%
\begin{equation}
K_{1}v_{i1x}\mathbf{=}M_{i1x}+K_{2}\mathbf{\nabla }\cdot \mathbf{v}_{i1}. 
\end{equation}%
We further apply the operator $K_{1}$ to Equation (A5) and use Equation
(A7). As a result, we find equation for $\mathbf{\nabla }\cdot \mathbf{v}%
_{i1}$%
\begin{eqnarray}
0 &=&K_{1}D_{ti}\left[ \frac{\partial }{\partial z}G_{i1z}+\frac{\partial }{%
\partial y}\left( G_{i1y}-C_{i1y}\right) \right] +\left[ D_{ti}\left( D_{ti}%
\frac{\partial }{\partial x}-\omega _{ci}\frac{\partial }{\partial y}\right) 
\mathbf{-}g_{i}\left( \frac{\partial ^{2}}{\partial y^{2}}+\frac{\partial
^{2}}{\partial z^{2}}\right) \right] M_{i1x}  \\
&&+\left\{ \left[ D_{ti}\left( D_{ti}\frac{\partial }{\partial x}-\omega
_{ci}\frac{\partial }{\partial y}\right) \mathbf{-}g_{i}\left( \frac{%
\partial ^{2}}{\partial y^{2}}+\frac{\partial ^{2}}{\partial z^{2}}\right) %
\right] K_{2}-K_{1}\left[ D_{ti}^{2}-c_{si}^{2}\left( \frac{\partial ^{2}}{%
\partial y^{2}}+\frac{\partial ^{2}}{\partial z^{2}}\right) \right] \right\} 
\mathbf{\nabla }\cdot \mathbf{v}_{i1}.  \nonumber
\end{eqnarray}

If we consider Equation (A8) without electromagnetic forces and the
background magnetic field, we will obtain equation%
\[
H_{i}\mathbf{\nabla }\cdot \mathbf{v}_{i1}=0, 
\]%
where 
\begin{eqnarray*}
H_{i} &=&-D_{ti}^{4}+D_{ti}^{2}c_{si}^{2}\left( \frac{\partial ^{2}}{%
\partial y^{2}}+\frac{\partial ^{2}}{\partial z^{2}}+\frac{\partial ^{2}}{%
\partial x^{2}}\right) +g_{i}\left[ \left( \gamma -1\right) g_{i}+\frac{%
\partial c_{si}^{2}}{\partial x}\right] \left( \frac{\partial ^{2}}{\partial
y^{2}}+\frac{\partial ^{2}}{\partial z^{2}}\right) \\
&&+D_{ti}^{2}\left( \frac{\partial c_{si}^{2}}{\partial x}-\gamma
g_{i}\right) \frac{\partial }{\partial x}.
\end{eqnarray*}%
This equation describe the ion sound and internal (ion) gravity waves. When
obtaining Equation (A8), we have excluded $v_{i1x}$. We also could exclude $%
\mathbf{\nabla }\cdot \mathbf{v}_{i1}$. In this case, the term $\partial
c_{si}^{2}/\partial x$ in the last term in the expression for $H_{i}$ would
have the sign $-$. It is connected with the form of dependence of $v_{i1x}$
from $\mathbf{\nabla }\cdot \mathbf{v}_{i1}$ and vice versa in Equations
(A4) and (A5).

Let us introduce the following notations:%
\begin{eqnarray}
N_{i1x} &\mathbf{=}&K_{1}D_{ti}\left[ \frac{\partial }{\partial z}G_{i1z}+%
\frac{\partial }{\partial y}\left( G_{i1y}-C_{i1y}\right) \right] +\left[
D_{ti}\left( D_{ti}\frac{\partial }{\partial x}-\omega _{ci}\frac{\partial }{%
\partial y}\right) \mathbf{-}g_{i}\left( \frac{\partial ^{2}}{\partial y^{2}}%
+\frac{\partial ^{2}}{\partial z^{2}}\right) \right] M_{i1x},  \\
K_{3} &=&\left\{ \left[ D_{ti}\left( D_{ti}\frac{\partial }{\partial x}%
-\omega _{ci}\frac{\partial }{\partial y}\right) \mathbf{-}g_{i}\left( \frac{%
\partial ^{2}}{\partial y^{2}}+\frac{\partial ^{2}}{\partial z^{2}}\right) %
\right] K_{2}-K_{1}\left[ D_{ti}^{2}-c_{si}^{2}\left( \frac{\partial ^{2}}{%
\partial y^{2}}+\frac{\partial ^{2}}{\partial z^{2}}\right) \right] \right\}
.  \nonumber
\end{eqnarray}%
Then Equation (A8) is given by 
\begin{equation}
K_{3}\mathbf{\nabla }\cdot \mathbf{v}_{i1}\mathbf{=}-N_{i1x}.  
\end{equation}%
Using notations (A6), we can represent the operator $K_{3}$ and the value $%
N_{i1x}$ defined by Equations (A9) in the form 
\begin{eqnarray}
K_{3} &=&\omega _{ci}^{2}D_{ti}\left( c_{si}^{2}\frac{\partial ^{2}}{%
\partial z^{2}}-D_{ti}^{2}\right) +\omega _{ci}D_{ti}^{2}\left[ \left(
\gamma -2\right) g_{i}+\frac{\partial c_{si}^{2}}{\partial x}\right] \frac{%
\partial }{\partial y}  \\
&&+D_{ti}^{3}\left[ c_{si}^{2}\left( \frac{\partial ^{2}}{\partial y^{2}}+%
\frac{\partial ^{2}}{\partial z^{2}}+\frac{\partial ^{2}}{\partial x^{2}}%
\right) -D_{ti}^{2}\right] +g_{i}D_{ti}\left[ \left( \gamma -1\right) g_{i}+%
\frac{\partial c_{si}^{2}}{\partial x}\right] \left( \frac{\partial ^{2}}{%
\partial y^{2}}+\frac{\partial ^{2}}{\partial z^{2}}\right)  \nonumber \\
&&+D_{ti}^{3}\left( -\gamma g_{i}+\frac{\partial c_{si}^{2}}{\partial x}%
\right) \frac{\partial }{\partial x}  \nonumber
\end{eqnarray}%
and%
\begin{eqnarray}
\frac{1}{D_{ti}}N_{i1x} &\mathbf{=}&D_{ti}\left[ D_{ti}\left( D_{ti}\frac{%
\partial }{\partial x}-\omega _{ci}\frac{\partial }{\partial y}\right) 
\mathbf{-}g_{i}\left( \frac{\partial ^{2}}{\partial y^{2}}+\frac{\partial
^{2}}{\partial z^{2}}\right) \right] G_{i1x}  \\
&&+\left[ D_{ti}^{2}\left( D_{ti}\frac{\partial }{\partial y}+\omega _{ci}%
\frac{\partial }{\partial x}\right) +g_{i}\left( D_{ti}\frac{\partial ^{2}}{%
\partial x\partial y}-\omega _{ci}\frac{\partial ^{2}}{\partial z^{2}}%
\right) \right] \left( G_{i1y}-C_{i1y}\right)  \nonumber \\
&&+\left[ D_{ti}\left( D_{ti}^{2}+\omega _{ci}^{2}\right) +g_{i}\left(
\omega _{ci}\frac{\partial }{\partial y}+D_{ti}\frac{\partial }{\partial x}%
\right) \right] \frac{\partial }{\partial z}G_{i1z}.  \nonumber
\end{eqnarray}

We further assume the following simplifications $\left( g_{i}\sim \partial
c_{si}^{2}/\partial x\right) $: 
\begin{eqnarray}
\omega _{ci}^{2} &\gg &c_{si}^{2}\left( \frac{\partial ^{2}}{\partial y^{2}}+%
\frac{\partial ^{2}}{\partial z^{2}}+\frac{\partial ^{2}}{\partial x^{2}}%
\right) ;D_{ti}^{2};g_{i}\frac{\partial }{\partial x},  \\
D_{ti} &\gg &\frac{g_{i}}{\omega _{ci}}\frac{\partial }{\partial y}%
,D_{ti}^{2}\gg \frac{g_{i}^{2}}{\omega _{ci}^{2}}\frac{\partial ^{2}}{%
\partial z^{2}}.  \nonumber
\end{eqnarray}%
Now, $D_{ti}=\partial /\partial t+v_{i0x}\partial /\partial x$. Then the
operator $K_{3}$ takes the simple form (see Eq. [A11]) 
\begin{equation}
K_{3}=\omega _{ci}^{2}D_{ti}\left( c_{si}^{2}\frac{\partial ^{2}}{\partial
z^{2}}-D_{ti}^{2}\right) +\omega _{ci}D_{ti}^{2}\left[ \left( \gamma
-2\right) g_{i}+\frac{\partial c_{si}^{2}}{\partial x}\right] \frac{\partial 
}{\partial y}.  
\end{equation}%
We take into account the last small term on the right hand-side of Equation
(A14) to obtain some additional terms proportional to $\omega _{ci}^{-2}$ in
expressions for components of $\mathbf{v}_{i1}$.

\subsection{The velocity $v_{i1x}$}

From Equations (A4), (A10), (A12), and (A14), we find the velocity $v_{i1x}$%
, using conditions (A13),%
\begin{eqnarray}
v_{i1x} &\mathbf{=}&\frac{D_{ti}}{\omega _{ci}^{2}}\frac{\left( c_{si}^{2}%
\frac{\partial ^{2}}{\partial y^{2}}+c_{si}^{2}\frac{\partial ^{2}}{\partial
z^{2}}-D_{ti}^{2}\right) }{\left( c_{si}^{2}\frac{\partial ^{2}}{\partial
z^{2}}-D_{ti}^{2}\right) }G_{i1x}\mathbf{+}\frac{1}{\omega _{ci}}\left(
G_{i1y}-C_{i1y}\right) \\
&&+\frac{D_{ti}}{\omega _{ci}^{2}}\frac{1}{\left( c_{si}^{2}\frac{\partial
^{2}}{\partial z^{2}}-D_{ti}^{2}\right) }\left( g_{i}-c_{si}^{2}\frac{%
\partial }{\partial x}\right) \frac{\partial }{\partial y}\left(
G_{i1y}-C_{i1y}\right) -\frac{c_{si}^{2}}{\omega _{ci}\left( c_{si}^{2}\frac{%
\partial ^{2}}{\partial z^{2}}-D_{ti}^{2}\right) }\frac{\partial ^{2}}{%
\partial y\partial z}G_{i1z}  \nonumber \\
&&-\frac{D_{ti}}{\omega _{ci}^{2}}\left[ \left( 1-\gamma \right)
g_{i}+c_{si}^{2}\frac{\partial }{\partial x}\right] \frac{1}{\left(
c_{si}^{2}\frac{\partial ^{2}}{\partial z^{2}}-D_{ti}^{2}\right) }\frac{%
\partial }{\partial z}G_{i1z}+\frac{D_{ti}}{\omega _{ci}^{2}}\frac{c_{si}^{2}%
\left[ \left( \gamma -2\right) g_{i}+\frac{\partial c_{si}^{2}}{\partial x}%
\right] }{\left( c_{si}^{2}\frac{\partial ^{2}}{\partial z^{2}}%
-D_{ti}^{2}\right) ^{2}}\frac{\partial ^{3}}{\partial y^{2}\partial z}%
G_{i1z}.  \nonumber
\end{eqnarray}%
When obtaining solution (A15), we have used some additional conditions%
\begin{eqnarray}
\left( c_{si}^{2}\frac{\partial ^{2}}{\partial y^{2}}+c_{si}^{2}\frac{%
\partial ^{2}}{\partial z^{2}}-D_{ti}^{2}\right) &\gg &\frac{D_{ti}}{\omega
_{ci}}c_{si}^{2}\frac{\partial ^{2}}{\partial x\partial y}, \\
\left( g_{i}-c_{si}^{2}\frac{\partial }{\partial x}\right) \frac{\partial }{%
\partial y} &\gg &\frac{g_{i}^{2}}{D_{ti}\omega _{ci}}\frac{\partial ^{2}}{%
\partial z^{2}};\frac{D_{ti}}{\omega _{ci}}c_{si}^{2}\left( \frac{\partial
^{2}}{\partial x^{2}}+\frac{\partial ^{2}}{\partial y^{2}}+\frac{\partial
^{2}}{\partial z^{2}}\right) ;\frac{D_{ti}^{3}}{\omega _{ci}},  \nonumber \\
D_{ti} &\gg &\frac{g_{i}}{\omega _{ci}}\left( \frac{\partial }{\partial y}%
\right) ^{-1}\left( \frac{\partial }{\partial z}\right) ^{2}.  \nonumber
\end{eqnarray}%
We note that in general expressions, we do not use the local approximation.

\bigskip

\subsection{The velocity $v_{i1y}$}

The velocity $v_{i1y}$, we find from Equation (A1). Applying the operator $%
D_{ti}$ to this equation and using Equations (23) and (24), we obtain 
\begin{equation}
D_{ti}^{2}v_{i1x}+g_{i}\frac{\partial v_{i1x}}{\partial x}\mathbf{=}%
D_{ti}G_{i1x}+D_{ti}\omega _{ci}v_{i1y}+\left[ g_{i}\left( 1-\gamma \right)
+c_{si}^{2}\frac{\partial }{\partial x}\right] \mathbf{\nabla }\cdot \mathbf{%
v}_{i1}.  
\end{equation}%
From Equations (A10), (A12), (A14), (A15), and (A17), we find, using
conditions (A13) and (A16),%
\begin{eqnarray}
v_{i1y} &\mathbf{=}&\mathbf{-}\frac{1}{\omega _{ci}}G_{i1x}-\frac{D_{ti}}{%
\omega _{ci}^{2}}\left[ \left( 1-\gamma \right) g_{i}+c_{si}^{2}\frac{%
\partial }{\partial x}\right] \frac{1}{\left( c_{si}^{2}\frac{\partial ^{2}}{%
\partial z^{2}}-D_{ti}^{2}\right) }\frac{\partial }{\partial y}G_{i1x} 
\\
&&\mathbf{+}\frac{D_{ti}}{\omega _{ci}^{2}}\left[ 1+\frac{c_{si}^{2}\frac{%
\partial ^{2}}{\partial x^{2}}}{\left( c_{si}^{2}\frac{\partial ^{2}}{%
\partial z^{2}}-D_{ti}^{2}\right) }\right] \left( G_{i1y}-C_{ie}\right) 
\nonumber \\
&&-\frac{D_{ti}}{\omega _{ci}^{2}}\frac{1}{\left( c_{si}^{2}\frac{\partial
^{2}}{\partial z^{2}}-D_{ti}^{2}\right) }\left[ \gamma g_{i}+\frac{c_{si}^{2}%
\frac{\partial ^{2}}{\partial z^{2}}}{\left( c_{si}^{2}\frac{\partial ^{2}}{%
\partial z^{2}}-D_{ti}^{2}\right) }\frac{\partial c_{si}^{2}}{\partial x}%
\right] \frac{\partial }{\partial x}\left( G_{i1y}-C_{ie}\right)  \nonumber
\\
&&+\frac{1}{\omega _{ci}^{2}D_{ti}}\frac{g_{i}}{\left( c_{si}^{2}\frac{%
\partial ^{2}}{\partial z^{2}}-D_{ti}^{2}\right) }\left[ \left( \gamma
-1\right) g_{i}+\frac{c_{si}^{2}\frac{\partial ^{2}}{\partial z^{2}}}{\left(
c_{si}^{2}\frac{\partial ^{2}}{\partial z^{2}}-D_{ti}^{2}\right) }\frac{%
\partial c_{si}^{2}}{\partial x}\right] \frac{\partial ^{2}}{\partial z^{2}}%
\left( G_{i1y}-C_{ie}\right)  \nonumber \\
&&+\frac{1}{\omega _{ci}}\left[ \left( 1-\gamma \right) g_{i}+c_{si}^{2}%
\frac{\partial }{\partial x}\right] \frac{1}{\left( c_{si}^{2}\frac{\partial
^{2}}{\partial z^{2}}-D_{ti}^{2}\right) }\frac{\partial }{\partial z}G_{i1z}
\nonumber \\
&&-\frac{1}{D_{ti}\omega _{ci}^{2}}\left\{ D_{ti}^{2}+\frac{g_{i}}{c_{si}^{2}%
}\left[ \left( \gamma -1\right) g_{i}+\frac{\partial c_{si}^{2}}{\partial x}%
\right] \right\} \frac{c_{si}^{2}}{\left( c_{si}^{2}\frac{\partial ^{2}}{%
\partial z^{2}}-D_{ti}^{2}\right) }\frac{\partial ^{2}}{\partial y\partial z}%
G_{i1z}  \nonumber \\
&&-\frac{D_{ti}}{\omega _{ci}^{2}}\left[ g_{i}\left( 1-\gamma \right)
+c_{si}^{2}\frac{\partial }{\partial x}\right] \frac{\left[ \left( \gamma
-2\right) g_{i}+\frac{\partial c_{si}^{2}}{\partial x}\right] }{\left(
c_{si}^{2}\frac{\partial ^{2}}{\partial z^{2}}-D_{ti}^{2}\right) ^{2}}\frac{%
\partial ^{2}}{\partial y\partial z}G_{i1z}. \nonumber
\end{eqnarray}

\bigskip

\subsection{The velocity $v_{i1z}$}

The velocity $v_{i1z}$, we find from Equation (A3). Applying the operator $%
D_{ti}$ and using Equation (24), we obtain 
\begin{equation}
D_{ti}^{2}v_{i1z}\mathbf{=}-g_{i}\frac{\partial }{\partial z}%
v_{i1x}+D_{ti}G_{i1z}+c_{si}^{2}\frac{\partial }{\partial z}\mathbf{\nabla }%
\cdot \mathbf{v}_{i1}.  
\end{equation}%
From Equations (A10), (A12), (A14), (A15), and (A19), we find, using
conditions given above,%
\begin{eqnarray}
v_{i1z} &\mathbf{=}&\frac{1}{\omega _{ci}}\frac{c_{si}^{2}}{\left( c_{si}^{2}%
\frac{\partial ^{2}}{\partial z^{2}}-D_{ti}^{2}\right) }\frac{\partial ^{2}}{%
\partial y\partial z}G_{i1x}+\frac{D_{ti}}{\omega _{ci}^{2}}\frac{1}{\left(
c_{si}^{2}\frac{\partial ^{2}}{\partial z^{2}}-D_{ti}^{2}\right) }\left(
g_{i}-c_{si}^{2}\frac{\partial }{\partial x}\right) \frac{\partial }{%
\partial z}G_{i1x} \\
&&-\frac{D_{ti}}{\omega _{ci}^{2}}\frac{c_{si}^{2}\left[ \left( \gamma
-2\right) g_{i}+\frac{\partial c_{si}^{2}}{\partial x}\right] }{\left(
c_{si}^{2}\frac{\partial ^{2}}{\partial z^{2}}-D_{ti}^{2}\right) ^{2}}\frac{%
\partial ^{3}}{\partial y^{2}\partial z}G_{i1x}+\frac{1}{\omega _{ci}}\frac{1%
}{\left( c_{si}^{2}\frac{\partial ^{2}}{\partial z^{2}}-D_{ti}^{2}\right) }%
\left( g_{i}-c_{si}^{2}\frac{\partial }{\partial x}\right) \frac{\partial }{%
\partial z}\left( G_{i1y}-C_{i1y}\right)  \nonumber \\
&&-\frac{1}{\omega _{ci}^{2}D_{ti}}\frac{c_{si}^{2}}{\left( c_{si}^{2}\frac{%
\partial ^{2}}{\partial z^{2}}-D_{ti}^{2}\right) }\left[ D_{ti}^{2}+\omega
_{bi}^{2}\frac{c_{si}^{2}\frac{\partial ^{2}}{\partial z^{2}}}{\left(
c_{si}^{2}\frac{\partial ^{2}}{\partial z^{2}}-D_{ti}^{2}\right) }-\frac{%
g_{i}^{2}}{c_{si}^{2}}\frac{D_{ti}^{2}}{\left( c_{si}^{2}\frac{\partial ^{2}%
}{\partial z^{2}}-D_{ti}^{2}\right) }\right] \frac{\partial ^{2}}{\partial
y\partial z}\left( G_{i1y}-C_{i1y}\right)  \nonumber \\
&&+\frac{D_{ti}}{\omega _{ci}^{2}}\frac{c_{si}^{2}\left[ \left( \gamma
-2\right) g_{i}+\frac{\partial c_{si}^{2}}{\partial x}\right] }{\left(
c_{si}^{2}\frac{\partial ^{2}}{\partial z^{2}}-D_{ti}^{2}\right) ^{2}}\frac{%
\partial ^{3}}{\partial x\partial y\partial z}\left( G_{i1y}-C_{i1y}\right)
\nonumber \\
&&-\frac{D_{ti}}{\left( c_{si}^{2}\frac{\partial ^{2}}{\partial z^{2}}%
-D_{ti}^{2}\right) }G_{i1z}+\frac{1}{\omega _{ci}}\frac{c_{si}^{2}\left[
\left( \gamma -2\right) g_{i}+\frac{\partial c_{si}^{2}}{\partial x}\right] 
}{\left( c_{si}^{2}\frac{\partial ^{2}}{\partial z^{2}}-D_{ti}^{2}\right)
^{2}}\frac{\partial ^{3}}{\partial y\partial z^{2}}G_{i1z}  \nonumber \\
&&-\frac{1}{\omega _{ci}^{2}D_{ti}}\frac{g_{i}}{\left( c_{si}^{2}\frac{%
\partial ^{2}}{\partial z^{2}}-D_{ti}^{2}\right) }\left[ \left( \gamma
-1\right) g_{i}+\frac{c_{si}^{2}\frac{\partial ^{2}}{\partial z^{2}}}{\left(
c_{si}^{2}\frac{\partial ^{2}}{\partial z^{2}}-D_{ti}^{2}\right) }\frac{%
\partial c_{si}^{2}}{\partial x}\right] \frac{\partial ^{2}}{\partial z^{2}}%
G_{i1z}.  \nonumber
\end{eqnarray}%
Note that the small term proportional to $\omega _{ci}^{-1}G_{i1z}$ in
Equation (A20) has appeared due to the small term in Equation (A14).

\bigskip
\begin{center}
\large{APPENDIX B}
\end{center}
\bigskip


\section{SOLUTION\ OF\ EQUATION\ (27)}

The components of Equation (27) are the following:%
\begin{equation}
0\mathbf{=}-\frac{1}{n_{e0}}\frac{\partial p_{e1}}{\partial x}-m_{i}g_{e}%
\frac{n_{e1}}{n_{e0}}+G_{e1x}+m_{e}\omega _{ce}v_{e1y},  
\end{equation}%
\begin{equation}
0\mathbf{=}-\frac{1}{n_{e0}}\frac{\partial p_{e1}}{\partial y}%
+G_{e1y}-C_{e1y}-m_{e}\omega _{ce}v_{e1x},  
\end{equation}%
\begin{equation}
0\mathbf{=}-\frac{1}{n_{e0}}\frac{\partial p_{e1}}{\partial z}+G_{e1z}. 
\end{equation}%
Below, we find the components of $\mathbf{v}_{e1}$.

\bigskip

\subsection{The velocity $v_{e1x}$}

The velocity $v_{e1x}$ can be easily found from Equations (B2) and (B3).
Differentiating Equation (B2) over $\partial /\partial z$ and (B3) over $%
\partial /\partial y$ and subtracting one equation from another, we obtain%
\begin{equation}
m_{e}\omega _{ce}\frac{\partial v_{e1x}}{\partial z}\mathbf{=}\frac{\partial 
}{\partial z}\left( G_{e1y}-C_{e1y}\right) -\frac{\partial }{\partial y}%
G_{e1z}.  
\end{equation}

\bigskip

\subsection{The value $\mathbf{\protect\nabla }\cdot \mathbf{v}_{e1}$}

We can find the value $\mathbf{\nabla }\cdot \mathbf{v}_{e1}$ from Equation
(B2). Applying to this equation operator $D_{te}$ and using Equations (36)
and (B4), we obtain%
\begin{eqnarray}
c_{se}^{2}\frac{\left( \gamma D_{te}+\Omega \right) }{\gamma \left(
D_{te}+\Omega \right) }\mathbf{\nabla }\cdot \mathbf{v}_{e1} &\mathbf{=}&%
\left[ g_{e}+\frac{\Omega }{\left( D_{te}+\Omega \right) }\frac{\partial
T_{e0}}{m_{i}\partial x}\right] v_{e1x}  \\
&&-\frac{\partial T_{e0}}{m_{i}\partial x}\frac{1}{B_{0}}\frac{D_{te}\Omega 
}{\left( D_{te}+\Omega \right) }\left( \frac{\partial }{\partial z}\right)
^{-1}B_{1x}\mathbf{-}\frac{1}{m_{i}}D_{te}\left( \frac{\partial }{\partial z}%
\right) ^{-1}G_{e1z}.  \nonumber
\end{eqnarray}

\bigskip

\subsection{The velocity $v_{e1y}$}

We calculate the velocity $v_{e1y}$ from Equation (B1), applying the
operator $D_{te}$ and using Equations (28) and (36). We further insert in
the equation obtained the value $\mathbf{\nabla }\cdot \mathbf{v}_{e1}$. The
important point is to differentiate carefully the background electron number
density and pressure. Then we use Equation (B4) for $v_{e1x}$. As a result
of calculations, we obtain 
\begin{eqnarray}
m_{e}\omega _{ce}v_{e1y} &\mathbf{=}&-G_{e1x}+\frac{g_{e}}{c_{se}^{2}}\left[
\left( \gamma -1\right) g_{e}+\frac{\partial c_{se}^{2}}{\partial x}\right] 
\frac{\gamma }{\left( \gamma D_{te}+\Omega \right) }\frac{m_{i}}{m_{e}\omega
_{ce}}\left( G_{e1y}-C_{e1y}\right)  \\
&&+\frac{\partial }{\partial x}\left( \frac{\partial }{\partial z}\right)
^{-1}G_{e1z}-\frac{1}{c_{se}^{2}}\left[ \left( \gamma -1\right) g_{e}\frac{%
\gamma D_{te}}{\left( \gamma D_{te}+\Omega \right) }+\frac{\partial
c_{se}^{2}}{\partial x}\right] \left( \frac{\partial }{\partial z}\right)
^{-1}G_{e1z}  \nonumber \\
&&-\frac{g_{e}}{c_{se}^{2}}\left[ \left( \gamma -1\right) g_{e}+\frac{%
\partial c_{se}^{2}}{\partial x}\right] \frac{\gamma }{\left( \gamma
D_{te}+\Omega \right) }\frac{m_{i}}{m_{e}\omega _{ce}}\frac{\partial }{%
\partial y}\left( \frac{\partial }{\partial z}\right) ^{-1}G_{e1z}  \nonumber
\\
&&+m_{i}g_{e}\frac{\partial T_{e0}}{T_{e0}\partial x}\frac{1}{B_{0}}\frac{%
\Omega }{\left( \gamma D_{te}+\Omega \right) }\left( \frac{\partial }{%
\partial z}\right) ^{-1}B_{1x}.  \nonumber
\end{eqnarray}

\bigskip

\subsection{The velocity $v_{e1z}$}

We find this velocity from $\mathbf{\nabla }\cdot \mathbf{v}_{e1}$%
\begin{equation}
\frac{\partial v_{e1z}}{\partial z}=\mathbf{\nabla }\cdot \mathbf{v}_{e1}-%
\frac{\partial v_{e1x}}{\partial x}-\frac{\partial v_{e1y}}{\partial y}. 
\end{equation}%
Using solutions (B4)-(B6), we obtain from Equation (B7)%
\begin{eqnarray}
\frac{\partial v_{e1z}}{\partial z} &=&\frac{1}{m_{e}\omega _{ce}}\frac{%
\partial }{\partial y}G_{e1x}-\frac{1}{m_{e}\omega _{ce}}\frac{\partial }{%
\partial x}\left( G_{e1y}-C_{e1y}\right)  \\
&&+\frac{1}{c_{se1}^{2}}\left[ g_{e}+\frac{\Omega }{\left( D_{te}+\Omega
\right) }\frac{\partial T_{e0}}{m_{i}\partial x}\right] \frac{1}{m_{e}\omega
_{ce}}\left( G_{e1y}-C_{e1y}\right)  \nonumber \\
&&-m_{i}\frac{g_{e}}{c_{se}^{2}}\left[ \left( \gamma -1\right) g_{e}+\frac{%
\partial c_{se}^{2}}{\partial x}\right] \frac{\gamma }{\left( \gamma
D_{te}+\Omega \right) }\left( \frac{1}{m_{e}\omega _{ce}}\right) ^{2}\frac{%
\partial }{\partial y}\left( G_{e1y}-C_{e1y}\right)  \nonumber \\
&&\mathbf{-}\frac{1}{c_{se1}^{2}m_{i}}D_{te}\left( \frac{\partial }{\partial
z}\right) ^{-1}G_{e1z}-\frac{\partial T_{e0}}{c_{se1}^{2}m_{i}\partial x}%
\frac{1}{B_{0}}\frac{D_{te}\Omega }{\left( D_{te}+\Omega \right) }\left( 
\frac{\partial }{\partial z}\right) ^{-1}B_{1x}.  \nonumber
\end{eqnarray}%
The velocity $c_{se1}$ is defined by expression (49). When obtaining
Equation (B8), we have used condition $D_{te}\gg g_{e}k_{y}/\omega _{ci}$.
In this case, $D_{te}=\partial /\partial t+v_{e0x}\partial /\partial x$.

\end{appendix}
\end{document}